\pdfoutput=1

\documentclass[11pt,twoside,a4paper,cmspaper,final,collab]{cms-tdr}
\def\svnVersion{485872P}\def\cmsCernNoTag{CERN-EP-2018-169}\def\cmsCernDate{\today}\def\cmsMessage{Published in the Journal of High Energy Physics as \href{http://dx.doi.org/10.1007/JHEP11(2018)172}{\doi{10.1007/JHEP11(2018)172.}}}

\begin{document}\cmsNoteHeader{B2G-17-004}

\hyphenation{had-ron-i-za-tion}
\hyphenation{cal-or-i-me-ter}
\hyphenation{de-vices}
\RCS$Revision: 479203 $
\RCS$HeadURL: svn+ssh://svn.cern.ch/reps/tdr2/papers/B2G-17-004/trunk/B2G-17-004.tex $
\RCS$Id: B2G-17-004.tex 479203 2018-10-25 08:52:06Z zucchett $
\newlength\cmsFigWidth
\setlength\cmsFigWidth{0.7\textwidth}
\providecommand{\cmsLeft}{left\xspace}
\providecommand{\cmsRight}{right\xspace}

\newcommand{\X}{\ensuremath{\cmsSymbolFace{X}}\xspace}
\newcommand{\V}{\ensuremath{\cmsSymbolFace{V}}\xspace}
\newcommand{\A}{\ensuremath{\cmsSymbolFace{A}}\xspace}
\newcommand{\PVpr}{\ensuremath{\mathrm{V'}}}
\newcommand{\AtoZh}{\ensuremath{\A\to\Z\Ph}\xspace}
\newcommand{\Phtobb}{\ensuremath{\Ph\to\bbbar}\xspace}
\newcommand{\VH}{\ensuremath{\V\Ph}\xspace}
\newcommand{\VV}{\ensuremath{\V\V}\xspace}
\newcommand{\ST}{\ensuremath{\PQt\text{+X}}\xspace}
\newcommand{\Vjets}{\ensuremath{\V\text{+jets}}\xspace}
\newcommand{\mVH}{\ensuremath{m_{\V\Ph}}\xspace}
\newcommand{\mtVH}{\ensuremath{m_{\V\Ph}^\mathrm{T}}\xspace}
\newcommand{\mX}{\ensuremath{m_{\X}}\xspace}
\newcommand{\mVpr}{\ensuremath{m_{\PVpr}}\xspace}
\newcommand{\mA}{\ensuremath{m_{\A}}\xspace}
\newcommand{\mH}{\ensuremath{m_{\PH}}\xspace}
\newcommand{\mHpm}{\ensuremath{m_{\PH^\pm}}\xspace}
\newcommand{\mh}{\ensuremath{m_{\Ph}}\xspace}
\newcommand{\mj}{\ensuremath{m_{\mathrm{j}}}\xspace}
\newcommand{\nsub}{\ensuremath{\tau_{21}}\xspace}
\newcommand{\B}{\ensuremath{\mathcal{B}}}
\newcommand{\gV}{\ensuremath{g_\text{V}}\xspace}
\newcommand{\cH}{\ensuremath{c_\text{H}}\xspace}
\newcommand{\cF}{\ensuremath{c_\text{F}}\xspace}
\newcommand{\cosba}{\ensuremath{\cos(\beta-\alpha)}\xspace}
\newcommand{\qqpbar}{\ensuremath{\PQq\overline{\PQq'}}\xspace}

\newlength\cmsTabSkip\setlength{\cmsTabSkip}{1ex}

\cmsNoteHeader{B2G-17-004}
\title{Search for heavy resonances decaying into a vector boson and a Higgs boson in final states with charged leptons, neutrinos and {\cPqb} quarks at $\sqrt{s}=13\TeV$}

\date{\today}

\abstract{
A search for heavy resonances, decaying into the standard model vector bosons and the standard model Higgs boson, is presented. The final states considered contain a {\cPqb} quark-antiquark pair from the decay of the Higgs boson, along with electrons and muons and missing transverse momentum, due to undetected neutrinos, from the decay of the vector bosons. The mass spectra are used to search for a localized excess consistent with a resonant particle. The data sample corresponds to an integrated luminosity of 35.9\fbinv collected in 2016 by the CMS experiment at the CERN LHC from proton-proton collisions at a center-of-mass energy of 13\TeV. The data are found to be consistent with background expectations. Exclusion limits are set in the context of spin-0 two Higgs doublet models, some of which include the presence of dark matter. In the spin-1 heavy vector triplet framework, mass-degenerate \PWpr and \PZpr resonances with dominant couplings to the standard model gauge bosons are excluded below a mass of 2.9\TeV at $95\%$ confidence level.
}

\hypersetup{%
pdfauthor={CMS Collaboration},%
pdftitle={Search for heavy resonances decaying into a vector boson and a Higgs boson in final states with charged leptons, neutrinos and b quarks at sqrt(s)=13 TeV},%
pdfsubject={CMS},%
pdfkeywords={CMS, B2G, Higgs, diboson, semileptonic, monoHiggs}
}

\maketitle

\section{Introduction}

The discovery and measurement of the mass and quantum numbers of a Higgs boson at the CERN LHC~\cite{Aad:2015zhl,Khachatryan:2016vau,bib:Aad20121,bib:Chatrchyan201230,Chatrchyan:2013lba} is consistent with the standard model (SM) of particle physics.
However, the proximity of the Higgs boson mass of 125\GeV~\cite{Aad:2015zhl} to the electroweak (EW) scale indicates either a significant amount of fine tuning, which mitigates the large quantum corrections to the Higgs boson mass, or the presence of new heavy particles near the EW scale~\cite{Barbieri:1987fn}. The relation between these heavy particles and the EW and Higgs sectors of the SM suggests that the new resonances may decay with a significant branching fraction into an SM vector boson ({\PW} or \Z) and an SM Higgs boson~(\Ph).

\begin{figure}[!htb]\centering
  \includegraphics[width=0.32\textwidth]{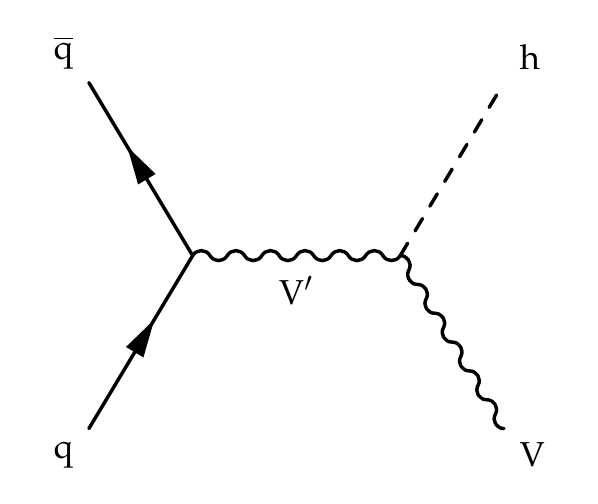}
  \includegraphics[width=0.32\textwidth]{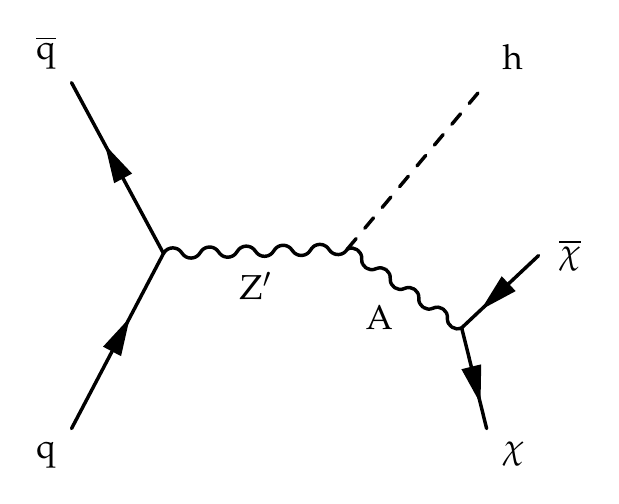}
  \\
  \includegraphics[width=0.32\textwidth]{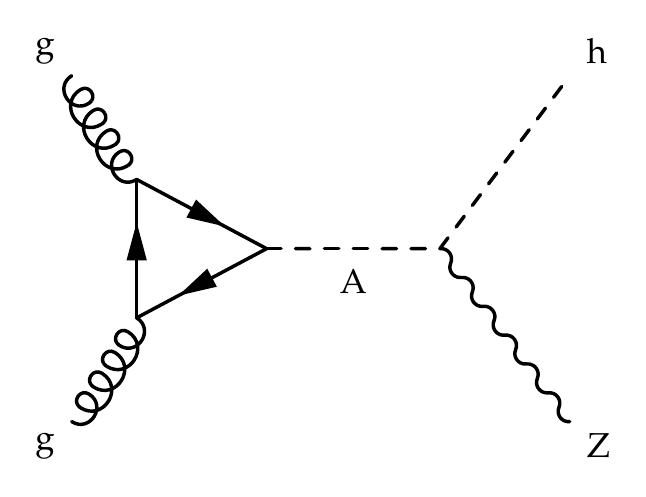}
  \includegraphics[width=0.32\textwidth]{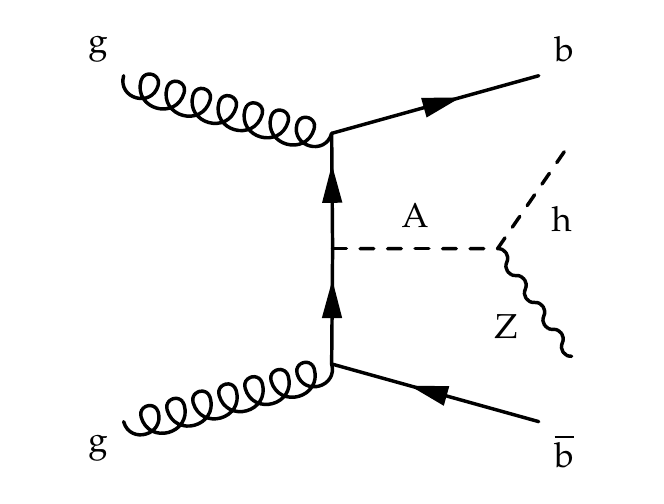}
  \caption{The leading order Feynman diagrams of the processes considered: heavy spin-1 vector boson production (\PVpr) and decay to an SM vector boson (\V) and a Higgs boson (\Ph) in the HVT framework (upper left); \PZpr boson that decays to a Higgs boson and an \A boson, with the latter decaying into dark matter particles ($\chi\overline{\chi}$), predicted by the \PZpr-2HDM model (upper right); production within the 2HDM model of a pseudoscalar \A boson through gluon-gluon fusion (lower left) and with accompanying {\cPqb} quarks (lower right).}
  \label{fig:Feynman}
\end{figure}

Several SM extensions containing extra $\mathrm{SU(2)}$ or $\mathrm{U(1)}$ gauge groups invoke massive gauge bosons (\PWpr and \PZpr) with weak couplings to the SM particles. Among these are the minimal \PWpr and \PZpr models, strongly coupled composite Higgs models, and little Higgs models~\cite{Grojean:2011vu,Barger:1980ix,1126-6708-2009-11-068,Contino2011,Marzocca2012,Bellazzini:2014yua,Lane:2016kvg,Han:2003wu,Schmaltz,Perelstein2007247}.
A large number of these models are described by the heavy vector triplet (HVT) framework~\cite{Pappadopulo2014}, which extends the SM by introducing a triplet of heavy vector bosons, one neutral ($\PZpr$) and two electrically charged ($\PW'^\pm$), which are degenerate in mass and are collectively referred to as \PVpr. The diagrams for these processes are depicted in Fig.~\ref{fig:Feynman} (upper left). In the HVT framework, $\gV$ is the coupling strength of the new interaction, $\cH$ is the coupling coefficient between the HVT bosons, the Higgs boson, and longitudinally polarized SM vector bosons, $\cF$ is the coupling coefficient between the HVT bosons and the SM fermions, and $g$ is the SM $\mathrm{SU(2)_L}$ gauge coupling. The coupling strength of the heavy vector bosons to SM bosons and fermions is determined by the $\gV\cH$ and $g^2 \cF / \gV$ parameters, respectively.
The HVT framework is presented in two scenarios, henceforth referred to as model~A and model~B, depending on the couplings to the SM particles~\cite{Pappadopulo2014}. In model~A ($\gV=1$, $\cH=-0.556$, $\cF=-1.316$), the coupling strengths to the SM bosons and fermions are comparable and the new particles decay primarily to fermions, as predicted by minimal $\PZpr$ and $\PWpr$ models.
In model~B ($\gV = 3$, $\cH=-0.976$, $\cF=1.024$), such as the composite Higgs models, the branching fraction to the SM bosons is nearly 100\% since the couplings to the SM fermions are small.

Heavy spin-0 resonances are also predicted in extensions of the SM Higgs sector, such as in two Higgs doublet models (2HDM)~\cite{bib:Branco20121}, which introduce a second scalar doublet in addition to the one from the SM. Different formulations of 2HDM predict different couplings of the two doublets to quarks and to massive leptons.
In Type-I 2HDM, all fermions couple to only one Higgs doublet, while in Type-II, the up- and down-type quarks couple to different doublets.
The two Higgs doublets entail the presence of five physical states: two neutral and CP-even bosons (\Ph and \PH, the latter being the more massive), a neutral and CP-odd boson (\A), and two charged scalar bosons ($\PH^\pm$). The dominant \A boson production process can be either through gluon-gluon fusion or through {\cPqb} quark associated production, as shown in Fig.~\ref{fig:Feynman} (lower), depending on the free parameters of the model, \tanb and $\alpha$, which are the ratio of the vacuum expectation values, and the mixing angle of the two Higgs doublets, respectively. In both cases, the heavy pseudoscalar boson \A may decay with a large branching fraction to a pair of \Z and Higgs bosons~\cite{bib:Branco20121}.

A particular formulation of the 2HDM, denoted as the \PZpr-2HDM model~\cite{Berlin:2014cfa}, is obtained by extending the 2HDM with an additional $\mathrm{U(1)_{\PZpr}}$ symmetry group that postulates a heavy spin-1 \PZpr particle with gauge coupling $g_{\PZpr}$, and a candidate for dark matter (DM), denoted as $\chi$, which couples to the \A boson with coupling strength $g_\chi$. In the process considered in this search, the \PZpr boson is produced from \qqbar annihilation, and decays into a pseudoscalar \A boson and a light Higgs boson. The Higgs boson decays to a {\cPqb} quark-antiquark pair ($\bbbar$), and the \A boson decays into a pair of DM particles ($\chi\overline{\chi}$), which escape detection, making this signature kinematically indistinguishable from the $\PZpr\to\Z\Ph\to\nu\nu\bbbar$ signal. The Feynman diagrams for the different signal processes are depicted in Fig.~\ref{fig:Feynman} (upper right).

Previous ATLAS and CMS searches~\cite{Aad:2015yza,Khachatryan:2015ywa,Sirunyan:2017acf,Aaboud:2016okv,Khachatryan2017137,Aaboud:2017cxo,Aaboud:2016lwx,Sirunyan:2017wto,Aaboud:2017ahz,Khachatryan:2015lba,Khachatryan:2016yji,Sirunyan:2017hnk,Aaboud:2017yqz} indicate that, in the framework of the models considered, the mass of the new resonance should exceed 1\TeV. Hence, the \V and Higgs boson from the subsequent decay have a large Lorentz boost, and thus the $\Ph\to\bbbar$ is reconstructed using a single large-cone jet containing the collimated decay products of the two hadronized {\cPqb} quarks.

This paper describes a search for heavy resonances, denoted as \X, decaying into an SM Higgs boson and a vector boson ({\PW} or \Z). The Higgs boson is assumed to decay to a \bbbar pair with a branching fraction of $58\%$~\cite{deFlorian:2016spz}, and the vector boson to decay to final states containing 0, 1, or 2 charged leptons ($\Z\to\nu\nu$, $\PW\to\ell\nu$, $\Z\to\ell\ell$), where $\ell$ denotes an electron or a muon, including those originating from a $\tau$ lepton decay. In the \PZpr-2HDM model, the $\Z\to\nu\nu$ decay is replaced by the $\A\to\chi\overline{\chi}$ decay to DM particles. The signal should appear as a localized excess in the mass spectra above the SM \Vjets and \ttbar backgrounds. The range of resonance mass \mX considered extends from 0.8\TeV, the minimum value that yields a sufficiently boosted Higgs boson, up to 4\TeV. 

This search is complementary to the CMS analysis targeting hadronic vector boson decays~\cite{Sirunyan:2017wto}, which excludes HVT triplets up to 3.1 and 3.3\TeV in models~A and~B, respectively, and retains a better sensitivity especially at low \mX thanks to the leptonic vector boson decays. The result of the present search significantly extends the sensitivity of the CMS searches in the same final state performed with 2.2--2.5\fbinv of data collected during 2015, which excluded a \PVpr ~boson with mass below 2.0\TeV in the HVT~model~B~\cite{Khachatryan2017137}, and a $m_{\PZpr} < 1.8\TeV$ and $\mA < 500\GeV$ in the \PZpr-2HDM model~\cite{Sirunyan:2017hnk}.

\section{The CMS detector}\label{sec:detector}

A detailed description of the CMS detector, together with a definition of the coordinate system used and the relevant kinematic variables, can be found in Ref.~\cite{Chatrchyan:2008zzk}.

The central feature of the CMS apparatus is a superconducting solenoid of 6\unit{m} internal diameter, providing a magnetic field of 3.8\unit{T}. Within the solenoid volume are a silicon pixel and strip tracker, a lead tungstate crystal electromagnetic calorimeter (ECAL), and a brass and scintillator hadron calorimeter (HCAL), each composed of a barrel and two endcap sections. Forward calorimeters extend the pseudorapidity ($\eta$) coverage provided by the barrel and endcap detectors. Muons are detected in gas-ionization chambers embedded in the steel flux-return yoke outside the solenoid.

The silicon tracker measures charged particles with $\abs{\eta} < 2.5$. It consists of 1440 silicon pixel and 15\,148 silicon strip detector modules. For nonisolated particles with transverse momenta of $1 < \pt < 10\GeV$ and $\abs{\eta} < 1.4$, the track resolutions are typically 1.5\% in \pt and 25--90 (45--150)\mum in the transverse (longitudinal) impact parameters~\cite{Chatrchyan:2014fea}.
The ECAL provides coverage up to $\abs{\eta} < 3.0$, and the energy resolution for unconverted or late-converting electrons and photons in the barrel section is about 1\% for particles that have energies in the range of tens of GeV. The dielectron mass resolution for $\Z \to\Pe \Pe$ decays when both electrons are in the ECAL barrel is 1.9\%, and is 2.9\% when both electrons are in the endcaps.
The HCAL covers the range of $\abs{\eta} < 3.0$, which is extended to $\abs{\eta} < 5.2$ through forward calorimetry.
The muon detectors, covering the range $\abs{\eta}< 2.4$, make use of three different technologies: drift tubes, cathode strip chambers, and resistive-plate chambers. The muon \pt resolution, as measured from tracks combining information from the silicon tracker and the muon detectors, is 2--10\% for muons with $0.1 < \pt < 1\TeV$~\cite{Chatrchyan:2012xi}.

The first level of the CMS trigger system~\cite{Khachatryan:2016bia}, composed of custom hardware processors, uses information from the calorimeters and muon detectors to select the most interesting events in a fixed time interval of less than 4\mus reducing the event rate from 40\unit{MHz} to approximately 100\unit{kHz}.
The high-level trigger (HLT) processor farm decreases the event rate from around 100\unit{kHz} to about 1\unit{kHz}, before data storage.

\section{Data and simulated samples}\label{sec:mcsimulation}

The data sample analyzed in this search corresponds to an integrated luminosity of 35.9\fbinv, collected with the CMS detector at the LHC in $\Pp\Pp$~collisions at a center-of-mass energy of 13\TeV.

The spin-1 gauge bosons \PWpr and \PZpr are simulated at leading order (LO) using the \MGvATNLO v2.4.2 matrix element generator~\cite{MadGraph}.
Different \mX hypotheses in the range of 800 to 4500\GeV are considered, assuming a resonance width narrow enough (0.1\% of the resonance mass) to be negligible compared to the experimental resolution, which is of the order of 4\%. This assumption is valid in a large fraction of the HVT parameter space, and fulfilled in both benchmark models~A and B~\cite{Pappadopulo2014}.
The \PWpr and \PZpr bosons decay to a Higgs boson and an SM boson ({\PW} or \Z); the former is required to decay into a \bbbar pair, and the SM vector bosons to electrons, muons, $\tau$ leptons, and neutrinos.

The spin-0 signal is generated at LO with \MGvATNLO in the gluon-gluon fusion and the {\cPqb} quark associated production processes separately, assuming a narrow resonance width. In the gluon-gluon fusion production mode, up to one additional jet is included in the final state, and only the top quark runs in the loop shown in Fig.~\ref{fig:Feynman}. The $\AtoZh$ decay is simulated with \textsc{MadSpin}~\cite{Artoisenet:2012st}.

The \PZpr-2HDM signal is generated at LO with \MGvATNLO assuming $g_{\PZpr}=0.8$, a unitary coupling of the \A boson to the DM candidate ($g_\chi = 1$), $\tan \beta = 1$, and mass-degenerate heavy Higgs bosons~\cite{Abercrombie:2015wmb}. In the case where $\cos(\beta -\alpha) \to 0$, also known as the alignment limit, the light Higgs boson is virtually indistinguishable from the SM Higgs boson, and its branching fractions match those of the SM one. This signal is characterized by the masses $m_{\PZpr}$ and $m_\A$, while the mass of the DM candidate $m_\chi$ does not affect the kinematic distributions significantly if the \A boson is on-shell. The DM particle mass is therefore set to a fixed value $m_\chi=100\GeV$ while $m_{\PZpr}$ is varied between 800 and 4000\GeV, and \mA between 300 and 800\GeV~\cite{Abercrombie:2015wmb}.

The SM backgrounds in this search are dominated by the inclusive production of \Vjets, with $\Z\to\nu\nu$, $\PW\to\ell\nu$, $\Z\to\ell\ell$, and \ttbar. The \Vjets events are simulated at LO with \MGvATNLO including up to 4 partons and normalized to the next-to-next-to-leading order (NNLO) cross section, computed using {\FEWZ} v3.1~\cite{Li:2012wna}. The \V boson \pt spectra are corrected to account for next-to-leading order (NLO) quantum chromodynamics (QCD) and EW contributions~\cite{Kallweit:2015dum}. Top quark pair (\ttbar) and single top quark $t$-channel and $\cPqt\PW$ productions are simulated at NLO with the \POWHEG v2 generator~\cite{Nason:2004rx,Frixione:2007vw,Alioli:2010xd}. The top quark pair production is rescaled to the cross section computed with \textsc{Top++} v2.0~\cite{Czakon:2011xx} at NNLO, and the transverse momenta of the top and antitop quarks are corrected to match the distribution observed in data~\cite{Khachatryan:2016mnb}. Other SM processes, such as \VV and \VH production, and single top quark (\ST) production in the $s$-channel, are simulated at NLO in QCD with \MGvATNLO using the {\scshape FxFx} merging scheme~\cite{Frederix:2012ps}. Events composed uniquely of jets arising from the SM strong interaction (QCD multijets) represent a minor background in the considered final states, and are estimated using LO samples produced with the same generator.

For all simulated samples, the hard scattering process uses the NNPDF 3.0~\cite{Ball:2014uwa} parton distribution functions (PDFs), and the generator is interfaced with \PYTHIA 8.205~\cite{Sjostrand:2007gs,Sjostrand:2006za} for the parton showering and hadronization. The CUETP8M1 underlying event tune~\cite{Skands:2014pea,Khachatryan:2015pea} is used in all samples, except for top quark pair production which is generated with the CUETP8M2T4 tune~\cite{CMS-PAS-TOP-16-021}.

Additional $\Pp\Pp$ interactions within the same or neighboring bunch crossings (pileup) are superimposed on the simulated processes, and the frequency distribution of the additional events is weighted to match the number of interactions per bunch crossing that was observed in 2016 data. Generated events are processed through a full CMS detector simulation based on {\GEANTfour}~\cite{Agostinelli:2002hh} and reconstructed with the same algorithms used for collision data.

\section{Event reconstruction}\label{sec:eventreconstruction}

A global event reconstruction is performed using a particle-flow (PF) algorithm~\cite{Sirunyan:2017ulk}, which uses an optimized combination of information from the various elements of the CMS detector to identify stable particles reconstructed in the detector as electrons, muons, photons, and charged or neutral hadrons.

Jets are reconstructed from PF candidates clustered using the anti-\kt algorithm~\cite{Cacciari:2008gp,Cacciari:2011ma} with a distance parameter $R=0.4$ (AK4 jets) or $R=0.8$ (AK8 jets).
The AK4 and AK8 jet four-momenta are obtained by clustering candidates passing the charged hadron subtraction (CHS) algorithm~\cite{CMS-PAS-JME-14-001}, which discards charged hadrons not originating from the primary vertex, by placing a restriction on the longitudinal impact parameter of the track.
The reconstructed vertex with the largest value of summed physics-object $\pt^2$ is taken to be the primary $\Pp\Pp$ interaction vertex. Here, the physics objects are the charged leptons, AK4 jets, and the associated missing transverse momentum \ptvecmiss, taken as the negative vector sum of the \pt of those jets.

The contribution of neutral particles originating from pileup interactions is proportional to the jet area and is estimated using the {\FASTJET} 3.0 package~\cite{Cacciari:2011ma,Cacciari:2008gn}, and then subtracted from the jet energy.
Jet energy corrections, estimated from simulation in dijet, multijet, $\gamma$+jets, and leptonically decaying Z+jets events, are applied as functions of the transverse momentum and pseudorapidity of the jet to correct the jet response. An adjustment is applied to account for residual differences between data and simulation. Jets are retained if their \pt exceeds 30\GeV for AK4 jets and 200\GeV for AK8 jets, and lie in the tracker acceptance $\abs{\eta}<2.4$. The jet energy resolution amounts typically to 5\% at 1\TeV~\cite{Khachatryan:2016kdb}.

The mass of the AK8 jet is measured after applying the pileup per particle identification (PUPPI) algorithm~\cite{Bertolini2014,CMS-PAS-JME-14-001}. The PUPPI algorithm uses a combination of the three-momenta of the particles, event pileup properties, and tracking information in order to compute a weight, assigned to charged and neutral PF candidates, describing the likelihood that each particle originates from a pileup interaction.
The weight for charged particles not coming from the primary vertex is 0, and it ranges from 0 to 1 for neutral particles. The weight is used to rescale the particle four-momenta, avoiding the need for further jet-area based pileup corrections. Jets are reconstructed from the PUPPI candidates using the anti-\kt algorithm with $R=0.8$. These jets are groomed using the soft-drop algorithm~\cite{Dasgupta:2013ihk,Larkoski:2014wba} to remove contributions from soft radiation and additional interactions, with algorithm parameters chosen to be $\beta = 0$ and $z_\text{cut} = 0.1$.
Dedicated mass corrections, derived from simulation and data in a region enriched with $\ttbar$ events with merged $\PW(\qqpbar)$ decays, are applied to the jet mass in order to remove residual jet \pt dependence~\cite{CMS-PAS-JME-16-003,Sirunyan:2017wto}, and to match the jet mass scale and resolution observed in data. The measured soft-drop jet mass resolution is approximately 10\%.
The AK8 soft-drop jets are split into two subjets by reverting the last step of the clustering algorithm applied to the jet constituents.

The combined secondary vertex algorithm~\cite{Sirunyan:2017ezt} is used for the identification of jets that originate from {\cPqb} quarks ({\cPqb} tagging), and is applied to both AK4 jets and AK8 subjets.
The algorithm uses the tracks and secondary vertices associated with AK4 jets or AK8 subjets as inputs to a neural network to produce a discriminator with values between 0 and 1, with higher values indicating a higher probability for the (sub)jet to originate from a {\cPqb} quark.
Selections on the discriminator output are applied, corresponding to a {\cPqb}-jet tagging efficiency for AK4 jets of 85 or 50\%, and a misidentification rate in a sample of quark and gluon jets of about 10 or 0.1\%. The {\cPqb} tagging efficiency in simulation is corrected to account for small residual differences between data and simulation~\cite{Sirunyan:2017ezt}.

Electrons are reconstructed in the fiducial region $\abs{\eta}<2.5$ by matching the energy deposits in the ECAL with tracks reconstructed in the tracker~\cite{Khachatryan:2015hwa}. The electron identification is based on the distribution of energy deposited along the electron trajectory, the direction and momentum of the track, and its compatibility with the primary vertex of the event.
Electrons are further required to be isolated from other energy deposits in the detector by applying an upper threshold on the isolation parameter.
The electron isolation parameter is defined as the sum of transverse momenta of all the PF candidates within $\Delta R = \sqrt{\smash[b]{(\Delta\eta)^2+(\Delta\phi)^2}} = 0.3$ around the electron direction, where $\phi$ is the azimuthal angle in radians, after the contributions from the electron itself, pileup and other reconstructed electrons are removed~\cite{Khachatryan:2015hwa}.

Muons are reconstructed within the acceptance of the CMS muon systems, $\abs{\eta}<2.4$, using the information from both the muon spectrometer and the silicon tracker~\cite{Chatrchyan:2012xi}. Muon candidates are identified via selection criteria based on the compatibility of tracks reconstructed from silicon tracker information only with tracks reconstructed from a combination of the hits in both the tracker and muon detector.
Additional requirements are based on the compatibility of the trajectory with the primary vertex, and on the number of hits observed in the tracker and muon systems.
Muons are required to be isolated by imposing a limit on the sum of reconstructed tracks within a cone $\Delta R = 0.4$ around the muon direction, ignoring the muon itself and tracks attributed to other muons~\cite{Chatrchyan:2012xi}.

Hadronically decaying $\tau$ leptons are reconstructed by combining one or three charged particle PF candidates with up to two neutral pion candidates~\cite{Khachatryan:2015dfa}.

\section{Event selection}
\label{sec:evtsel}

Events are divided into categories depending on the number and flavor of the reconstructed charged leptons. The zero-lepton ($0\ell$), the single-lepton ($1\ell$), and double-lepton ($2\ell$) channels are separated according to the electron and muon content in the event. These channels have different selections, aiming at maximizing the $\PVpr$ signal significance. Events are further categorized depending on the number of {\cPqb}-tagged subjets (1 or 2) passing the 85\% efficient {\cPqb} tagging selection. In total, 10 exclusive categories are defined.

The identification criteria for the boosted $\Ph\to\bbbar$ candidate (\Ph jet) are the same for all event categories. The highest-\pt AK8 jet in the event is required to have $\pt>200\GeV$ and $\abs{\eta}<2.5$. Its soft-drop jet mass \mj must satisfy $105 < \mj < 135\GeV$ for the event to enter the signal region~(SR). In order to discriminate against the copious vector boson production in association with quark and gluon jets, and to retain the maximum signal efficiency over the whole of the \pt range of the \Ph jet, the \Ph jet is required to have 1 or 2 {\cPqb}-tagged subjets; otherwise the event is discarded. The 2 {\cPqb}-tagged subjet categories dominate the sensitivity at low \mX, but because of the decrease in efficiency of track reconstruction at very large jet \pt, and the overlap between the two subjets of the \Ph jet, at high \mX, a significant number of signal events is retained in the 1 {\cPqb}-tagged subjet categories.
The \Ph jet tagging efficiency ranges between 13 and 24\% in the 1 {\cPqb} tag categories, and 29 and 19\% in the 2 {\cPqb} tag categories, respectively, at low and high \mX. The average probability for a \Vjets event to pass the \Ph jet selections is 1.7 and 0.2\% in the 1 and 2 {\cPqb} tag categories; the mistag rate for \ttbar events is generally larger, and corresponds to 2.9 and 0.5\%, respectively.

In the $0\ell$ channel, signal events are expected to have a large \ptmiss, defined as the magnitude of \ptvecmiss, arising from the boosted \Z boson decaying into a pair of neutrinos or from the \A boson decaying to a pair of DM particles, which escape undetected. Data are collected using HLT algorithms that require a \ptmiss, calculated either with or without considering muons, or missing hadronic activity \mht~\cite{Khachatryan:2016bia} larger than 90--110\GeV, depending on the data taking period.
The reconstructed \ptmiss is required to be larger than 250\GeV to ensure that the trigger is fully efficient.
The multijet production is suppressed by requiring that the minimum azimuthal angular separations between all AK4 and AK8 jets and the missing transverse momentum vector satisfies $\Delta\phi\text{(jet, \ptvecmiss)} > 0.5$. The \Ph jet must fulfill a tighter requirement $\Delta\phi (\vec{\pt}^\Ph, \ptvecmiss) > 2.0$ and the fraction of its momentum given by the charged-hadron candidates has to be larger than $0.1$ to remove events arising from detector noise.
Events containing isolated leptons with $\pt>10\GeV$ or hadronically decaying $\tau$ leptons with $\pt>18\GeV$ are removed in order to reduce the contribution from other SM processes.
The \ttbar background contribution is reduced by removing events in which any additional AK4 jet not overlapping with the \Ph jet within $\Delta R (\text{jet},\Ph) > 0.8$ is {\cPqb} tagged using a selection which is 85\% efficient on genuine {\cPqb} jets.
Because of the lack of visible decay products from the $\Z\to\nu\nu$ and $\A\to\chi\chi$ bosons, direct reconstruction of the resonance mass is not possible. Instead, the Higgs boson jet momentum and the \ptvecmiss are used to compute the transverse mass  $\mtVH = \sqrt{\smash[b]{2 \ptmiss \pt^\Ph\, [1-\cos{\Delta \phi (\ptvecmiss, \vec{\pt}^\Ph) }]}}$.

Events in the $1\Pe$ channel are collected using a trigger requiring either an isolated electron with $\pt > 32\GeV$ or an electron with no isolation requirement and $\pt > 115\GeV$. The $1\mu$ channel requires at least one muon with $\pt > 50\GeV$ and no selection on isolation. In addition, the same set of triggers for the $0\ell$ channel is also used for the $1\ell$ channels to take advantage of the large \ptmiss and \mht from the escaping neutrino from the {\PW} boson decay.
Offline, events are retained if exactly one lepton satisfies a \pt threshold of $55\GeV$ and the electron and muon identification and isolation selections.
The efficiencies of these selection criteria are approximately 75 and 95\%, respectively. Correction factors are applied to account for small differences between data and simulation in the trigger selection, and lepton reconstruction, identification and isolation. In the $1\Pe$ channel, the multijet background is further suppressed by requiring $\ptmiss>80\GeV$. Azimuthal angular separations are imposed, $\Delta \phi (\ell, \ptvecmiss) < 1.5$, $\Delta \phi (\ell, \Ph) > 2.0$, and $\Delta \phi (\vec{\pt}^\Ph, \ptvecmiss) > 2.0$ to select a topology where the vector boson recoils against the Higgs boson jet.
As in the $0\ell$ selection, events with additional {\cPqb}-tagged AK4 jets are vetoed to reduce the \ttbar background contamination.
The four-momentum of the neutrino is estimated using a kinematic reconstruction technique~\cite{Khachatryan2017137}. The $p_x^\nu$ and $p_y^\nu$ components of the neutrino momentum in the transverse plane are assumed to be equal to the ones of \ptvecmiss. By constraining the invariant mass of the sum of the charged lepton and neutrino four-momenta to be consistent with the {\PW} boson mass, a quadratic equation is derived for the longitudinal component of the neutrino momentum, $p_z^\nu$. The reconstructed $p_z^\nu$ is chosen to be the real solution with the lower magnitude or, where both the solutions are complex, the real part of the solutions.
The sum of the neutrino and the lepton four-momenta is used to reconstruct the {\PW} boson candidate, and subsequently, in combination with the \Ph jet four-momentum, the resonance candidate mass \mVH. The reconstructed {\PW} boson candidate has to have a transverse momentum larger than 200\GeV and a pseudorapidity separation $\abs{ \Delta \eta (\PW, \Ph) } < 3$, otherwise the event is discarded.

The $2\ell$ channel accepts events collected with the same triggers as in the $1\ell$ channel. An additional isolated electron or muon is required to have $\pt > 20\GeV$ and the same flavor and opposite charge as the leading one. The identification and isolation requirements are looser than those in the $1\ell$ channel, and the selection efficiency does not strongly depend on $\Delta R(\ell\ell)$ and is between 85 and 90\% for the electron pair, and 90 and 95\% for the muon pair.
The leptonic \Z boson candidates require the dilepton invariant mass to be between $70$ and $110\GeV$, and the transverse momentum to be greater than 200\GeV. Additionally, the separation in $\eta$ between the \Z boson candidate and the Higgs boson jet is required to satisfy $\abs{\Delta \eta (\Z, \Ph)} < 1.3$ and $\Delta\phi(\Z, \Ph)<2.0$ to partially reduce the dominant Z+jets background and increase the signal significance at low \mX, where the $2\ell$ channel adds most to the sensitivity. Since the \ttbar contribution is small, no veto on additional {\cPqb}-tagged AK4 jets is applied. The resonance candidate mass \mVH is defined as the invariant mass of the \Z boson and the \Ph jet.

\begin{figure}[!htb]\centering
  \includegraphics[width=0.7\textwidth]{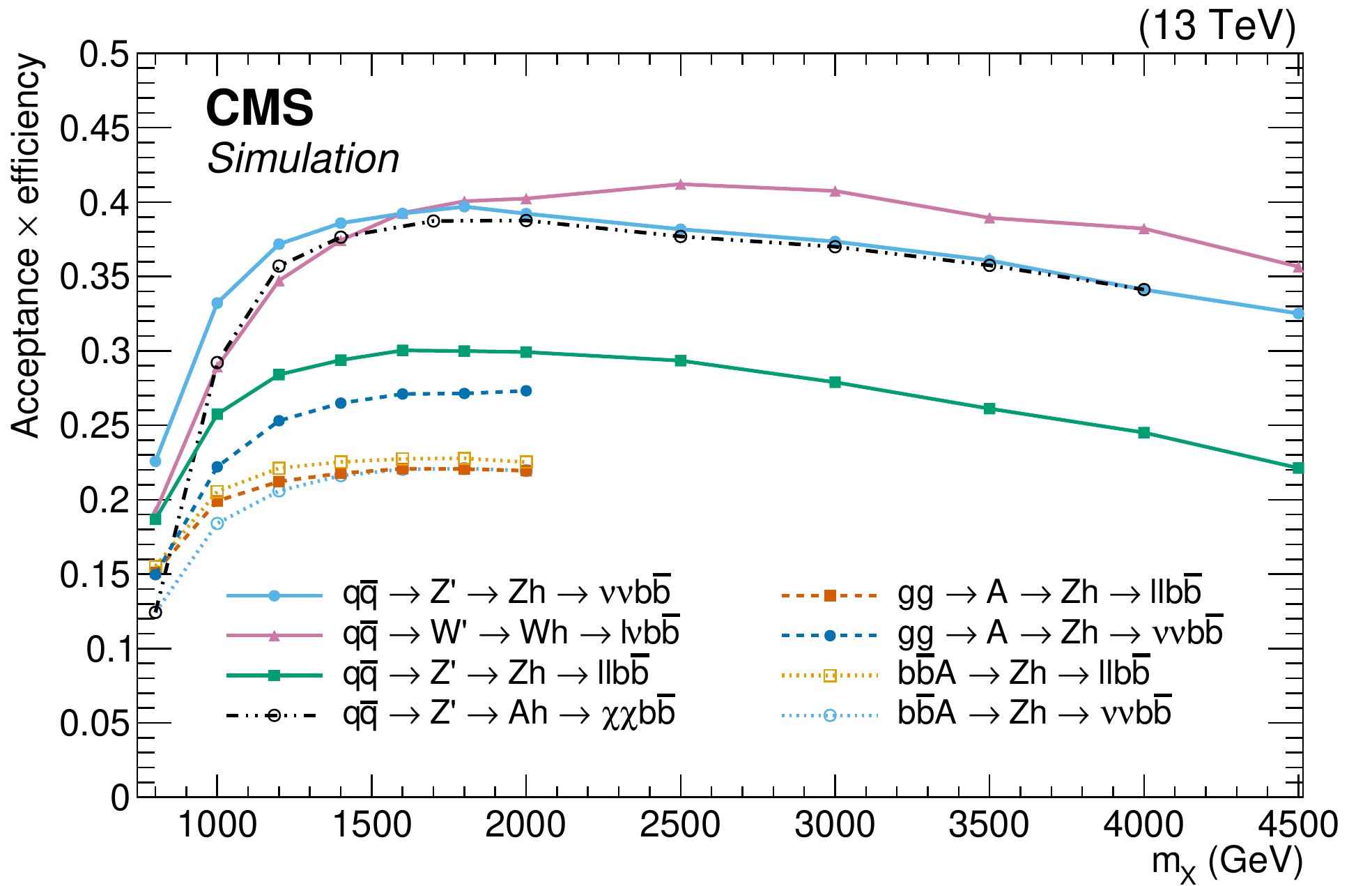}
  \caption{The product of acceptance and efficiency for the various signal processes and for different assumed masses of the resonances $m_{\PVpr}$ or \mA. The dash-dotted and solid lines indicate spin-0 and spin-1 resonances, respectively, in different production or decay modes. The dashed line represents the spin-1 resonance in the \PZpr-2HDM model with $\mA=300\GeV$. The efficiencies are derived by considering only the relevant decay modes of the vector bosons ($\Pe$, $\mu$, or $\tau$), and represent the sum of the efficiencies in the 1 and 2 {\cPqb} tag categories.}
  \label{fig:Efficiency}
\end{figure}

A further requirement, applied in all channels, is to have either \mtVH or \mVH larger than 750\GeV, in order to ensure a sufficiently large Lorentz boost for the Higgs boson.
The average signal acceptance times efficiency, derived taking into account the leptonic branching fractions with respect to the leptonic decay modes of the vector bosons ($\nu$ or $\Pe$, $\mu$, and $\tau$) and summing the 1 and 2 {\cPqb} tag categories, is shown in Fig.~\ref{fig:Efficiency} for the different signal models.

\section{Background estimation}
\label{sec:bkg}

A signal would produce a narrow peak above a smoothly falling background in the distribution of the kinematic variables \mVH or \mtVH.
The main background consists of a leptonically decaying vector boson in association with a jet from {\cPqb} or light-flavor quarks, or gluons, where the light quark or gluon jets are misidentified as {\cPqb} jets (\Vjets).
A sizable background originates from top quark events (\ttbar and \ST), whose contribution can be as large as 60\% in the $1\ell$ category. Minor contributions come from \VV, \VH, and multijet processes.
The \Vjets and \ttbar backgrounds are estimated using two different procedures based on data and simulation.

\subsection{Background normalization}

The normalization of the simulated top quark background is corrected with a scale factor determined in eight dedicated control regions, defined by inverting one selection criteria and removing the \mj requirement. In the $0\ell$, $1\Pe$ and $1\mu$ categories, the veto on additional {\cPqb}-tagged AK4 jets is inverted by requiring at least one additional AK4 jet passing the {\cPqb} tagging selection with a 0.1\% mistag rate to obtain a higher \ttbar purity. In the $2\ell$ categories, the leptons are required to have opposite sign and different flavor (one electron and one muon), and the two leptons must have $m_{\Pe\mu} > 110\GeV$ and $\pt^{\Pe\mu} > 120\GeV$ to give distributions similar to those in the SR.
After subtracting the remaining contribution of the other backgrounds, the scale factors are derived for each control region from the ratio of event yields between data and simulation. The scale factors are then applied to the simulated events in the corresponding SR; the scale factors derived in the $1\Pe,1\mu$ top quark control regions are used to correct the top quark yields in the $2\Pe$ and $2\mu$ categories. The top quark background scale factors are given in Table~\ref{tab:TopCR}.

\begin{table}[!htb]
  \topcaption{The scale factors (SF) derived to correct for the event yields of the \ttbar and \ST backgrounds in simulation for different top quark control regions. The uncertainties arising from the limited size of the data samples (stat.) and systematic effects (syst.), described in Section~\ref{sec:sys}, are reported.}\label{tab:TopCR}
  \centering
    \begin{tabular}{cc cccc}
      \hline
      \multicolumn{2}{c}{Control region} & $\ttbar, \ST$ SF $\pm$ stat. $\pm$ syst. \\
      \hline
      \multirow{4}{*}{1 {\cPqb} tag}
      & $0\ell$ & 1.02 $\pm$ 0.04 $\pm$ 0.25 \\
      & $1\Pe$ & 0.91 $\pm$ 0.02 $\pm$ 0.25 \\
      & $1\mu$ & 0.89 $\pm$ 0.02 $\pm$ 0.25 \\
      & $1\Pe,1\mu$ & 0.94 $\pm$ 0.06 $\pm$ 0.23 \\
      [\cmsTabSkip]
      \multirow{4}{*}{2 {\cPqb} tag}
      & $0\ell$ & 1.05 $\pm$ 0.10 $\pm$ 0.26 \\
      & $1\Pe$ & 0.94 $\pm$ 0.04 $\pm$ 0.26 \\
      & $1\mu$ & 0.85 $\pm$ 0.03 $\pm$ 0.26 \\
      & $1\Pe,1\mu$ & 1.03 $\pm$ 0.17 $\pm$ 0.23 \\
      \hline
    \end{tabular}
\end{table}

The \Vjets background prediction is performed through a two stage procedure based on data. In the first stage, the normalization is determined from a fit to the data in the \mj distribution. In the second stage, the \mVH and \mtVH distributions are estimated using the data in the \mj sidebands and a transfer function derived from simulation.

The \Vjets event yield in the SR is estimated through a parametrization of the \mj distributions, considering the three separate components \Vjets, \ttbar and \ST, and the sum of the SM diboson processes and the SM Higgs production processes. The latter contributes up to 50--70\% of the total SM diboson yield in the 2 {\cPqb}-tagged categories, and 6--10\% in the single {\cPqb}-tagged categories.
The \mj distributions are modeled using analytic functions, chosen based on studies in simulation. The \mj spectrum in \Vjets events consists of a falling distribution and is parametrized by a polynomial with 3--5 parameters depending on the signal event category. The \mj distribution from the top quark background, however, has two peaks, one corresponding to a Lorentz-boosted $\PW\to\qqpbar$ decay, and the other corresponding to the top quark mass in events where the top quark is sufficiently boosted for all $\mathrm{t}\to\PW\cPqb \to \qqpbar\cPqb$ decay products to be merged in a single AK8 jet.
The function describing the top quark mass spectrum is determined from simulation, and the normalization is constrained from the dedicated control regions, as given in Table~\ref{tab:TopCR}. Diboson samples present peaks corresponding to the {\PW}, \Z, and Higgs boson masses, and both the \mj distributions and their event yields are taken from simulation.

\begin{figure*}[!hbtp]\centering
  \includegraphics[width=0.40\textwidth]{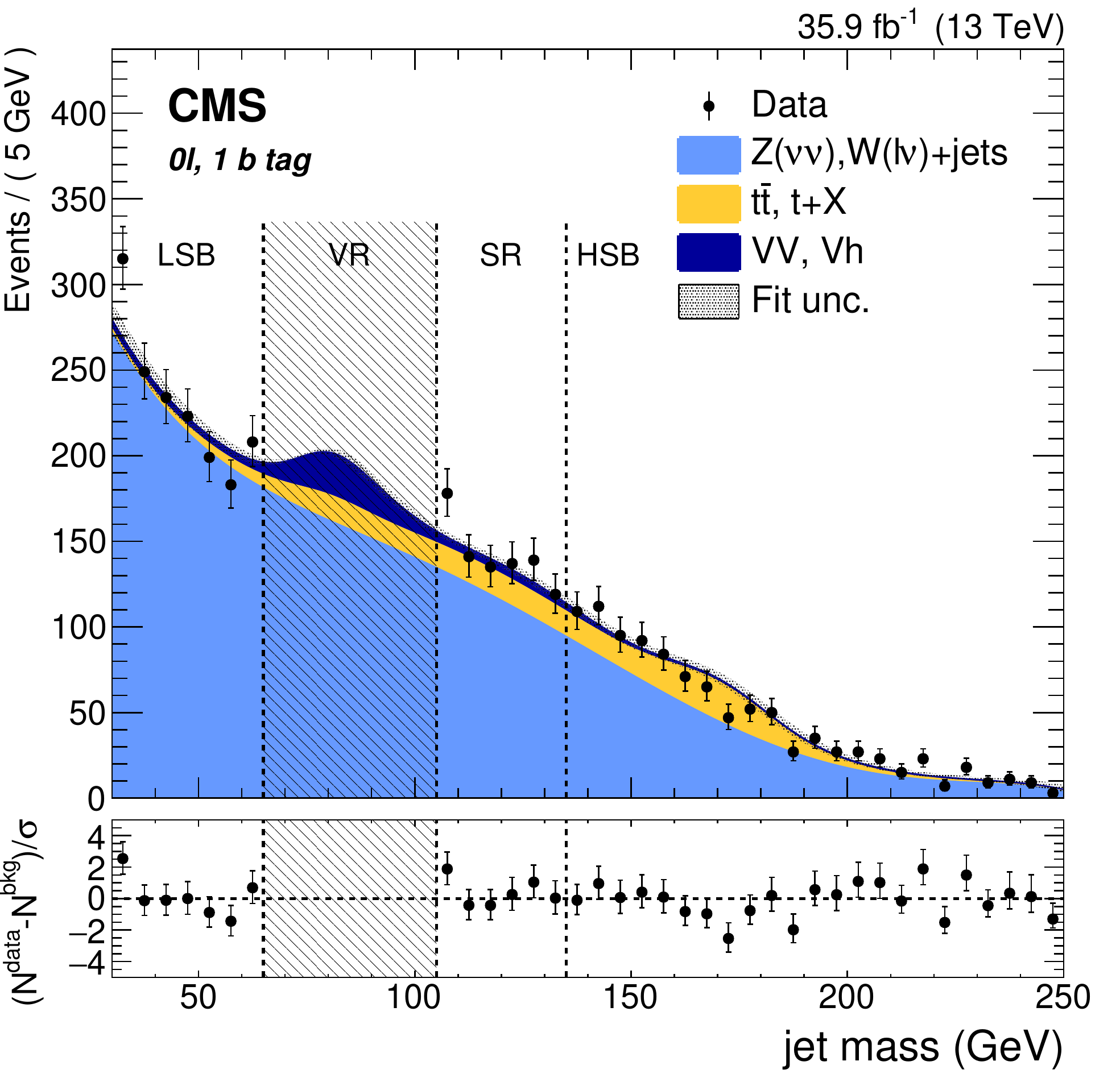}
  \includegraphics[width=0.40\textwidth]{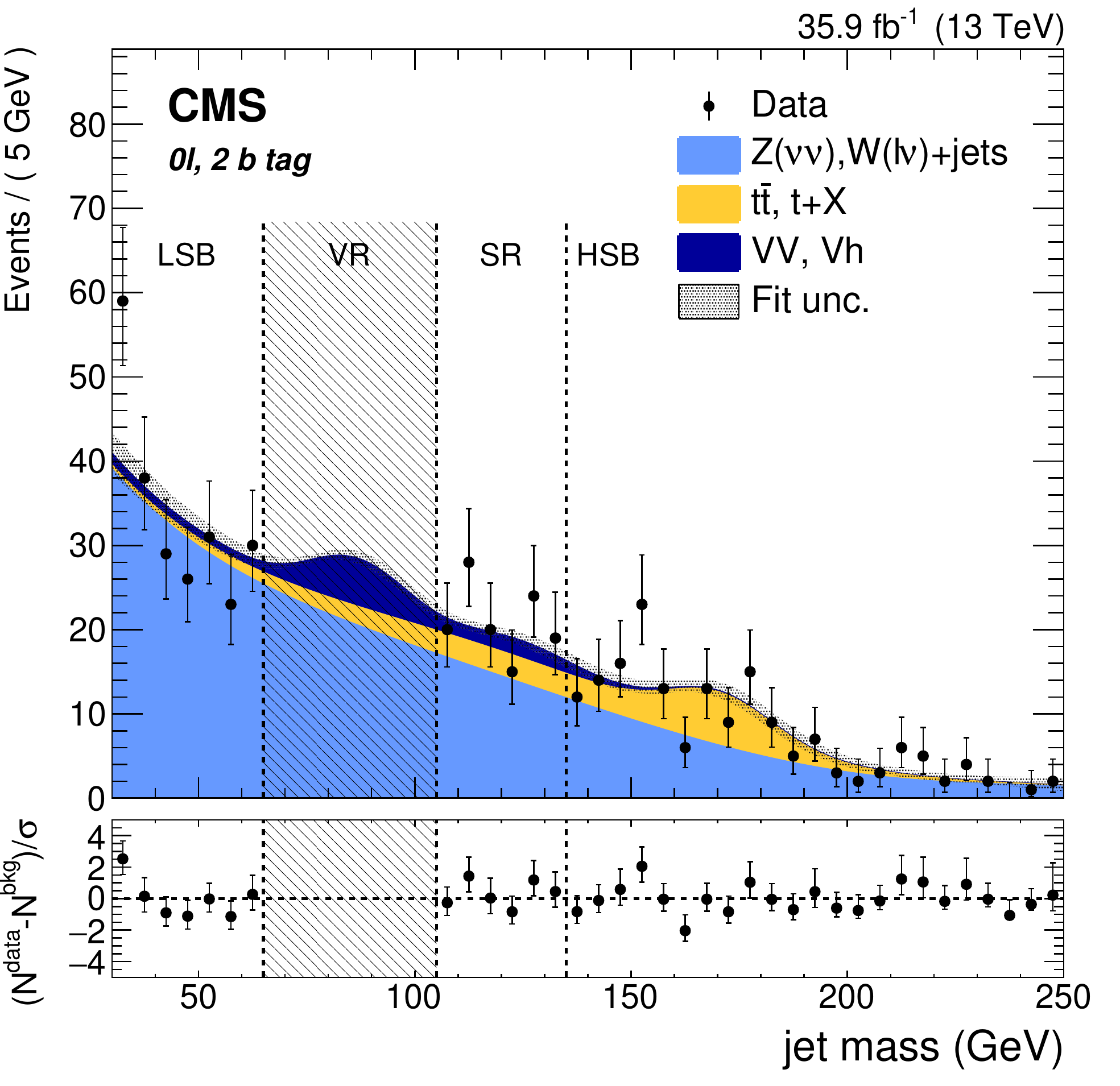}

  \includegraphics[width=0.40\textwidth]{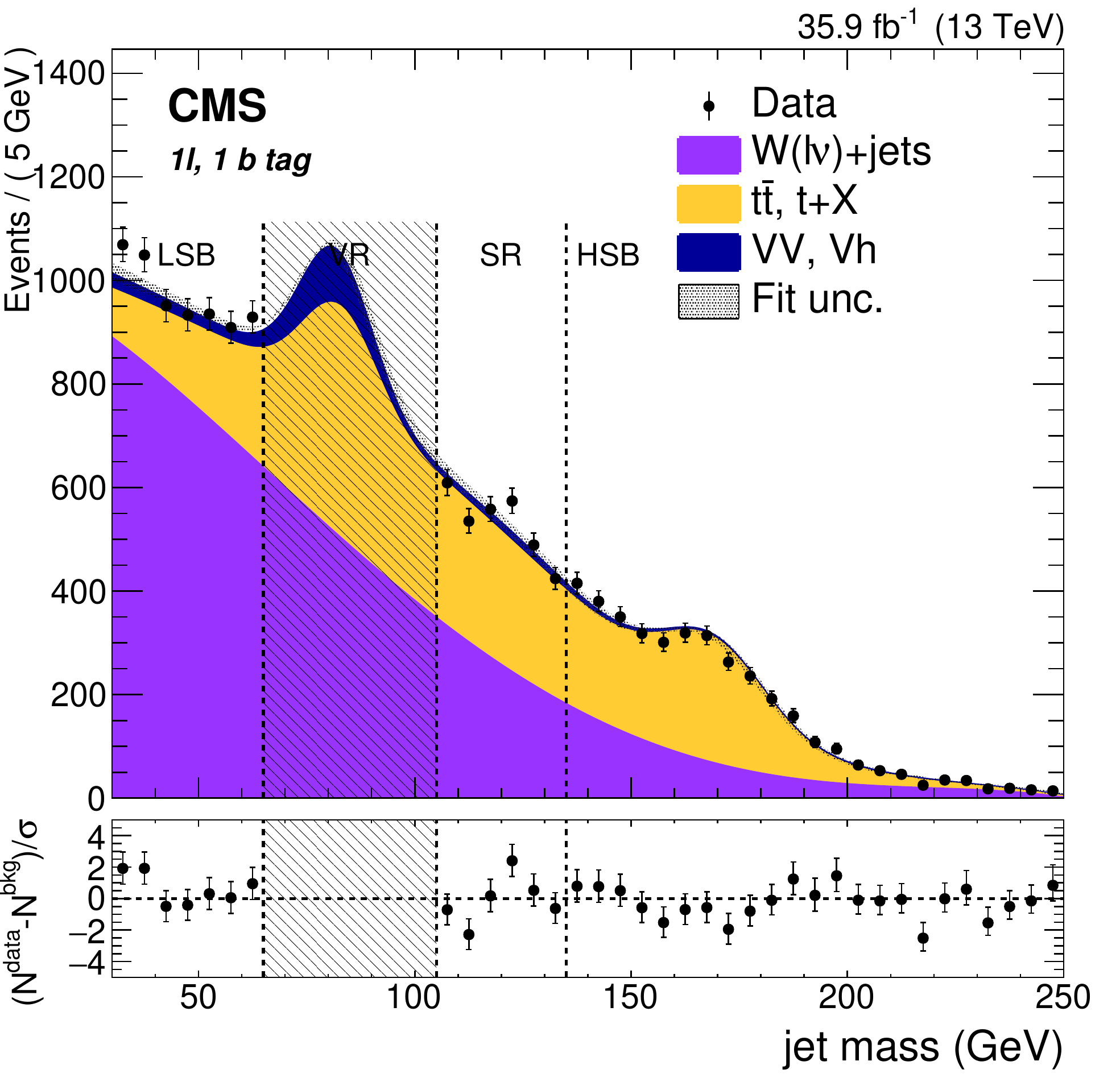}
  \includegraphics[width=0.40\textwidth]{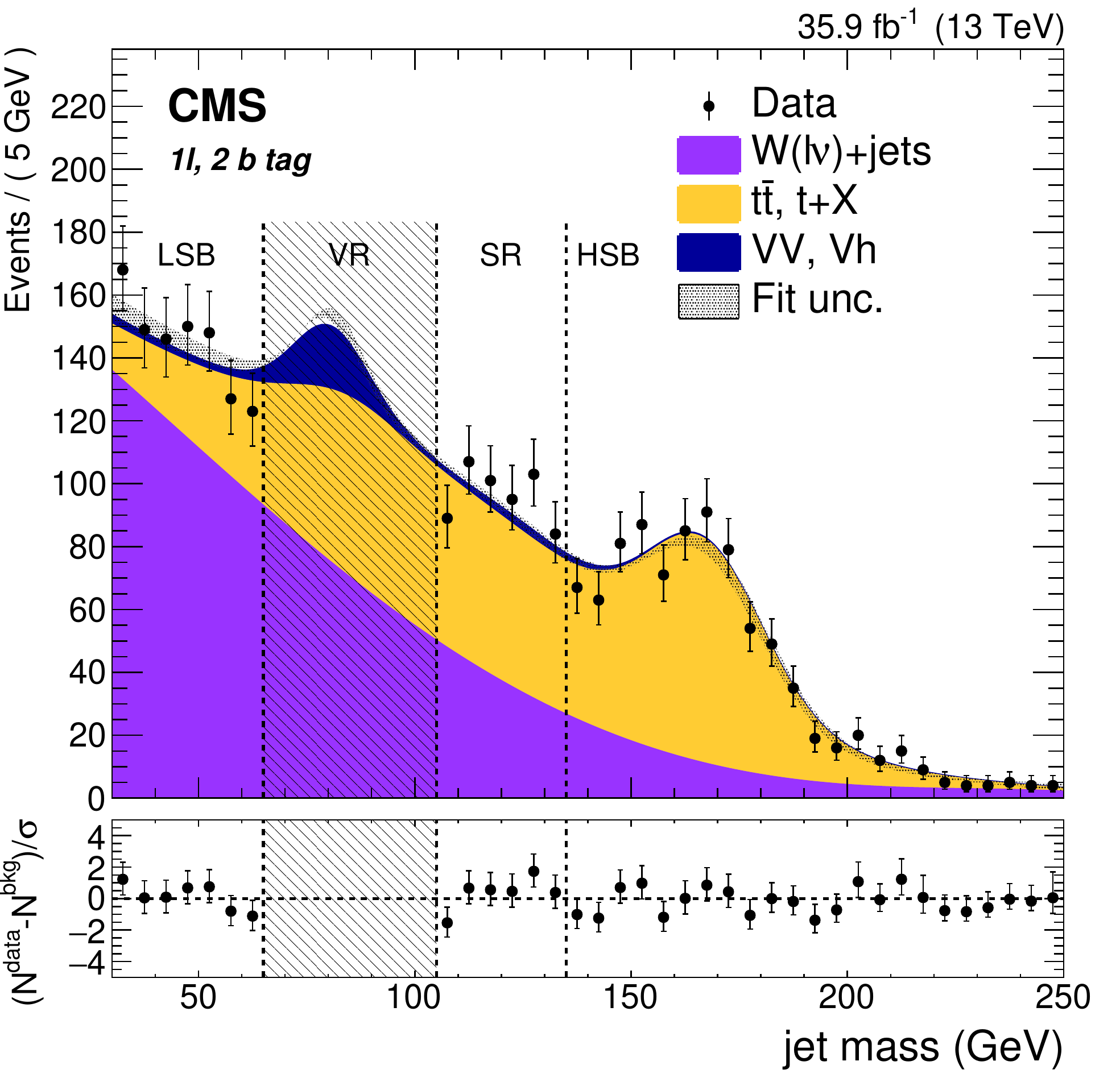}

  \includegraphics[width=0.40\textwidth]{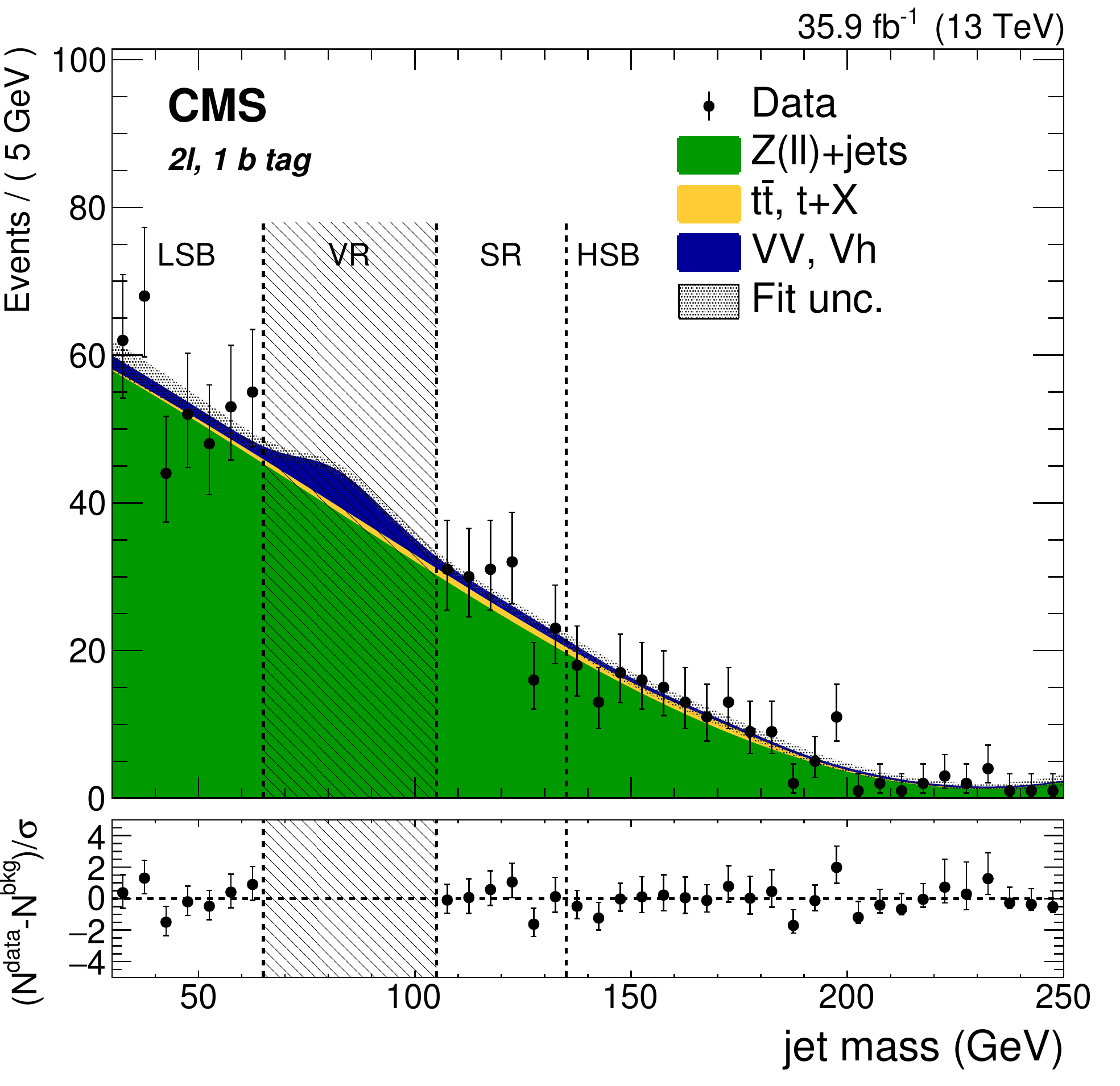}
  \includegraphics[width=0.40\textwidth]{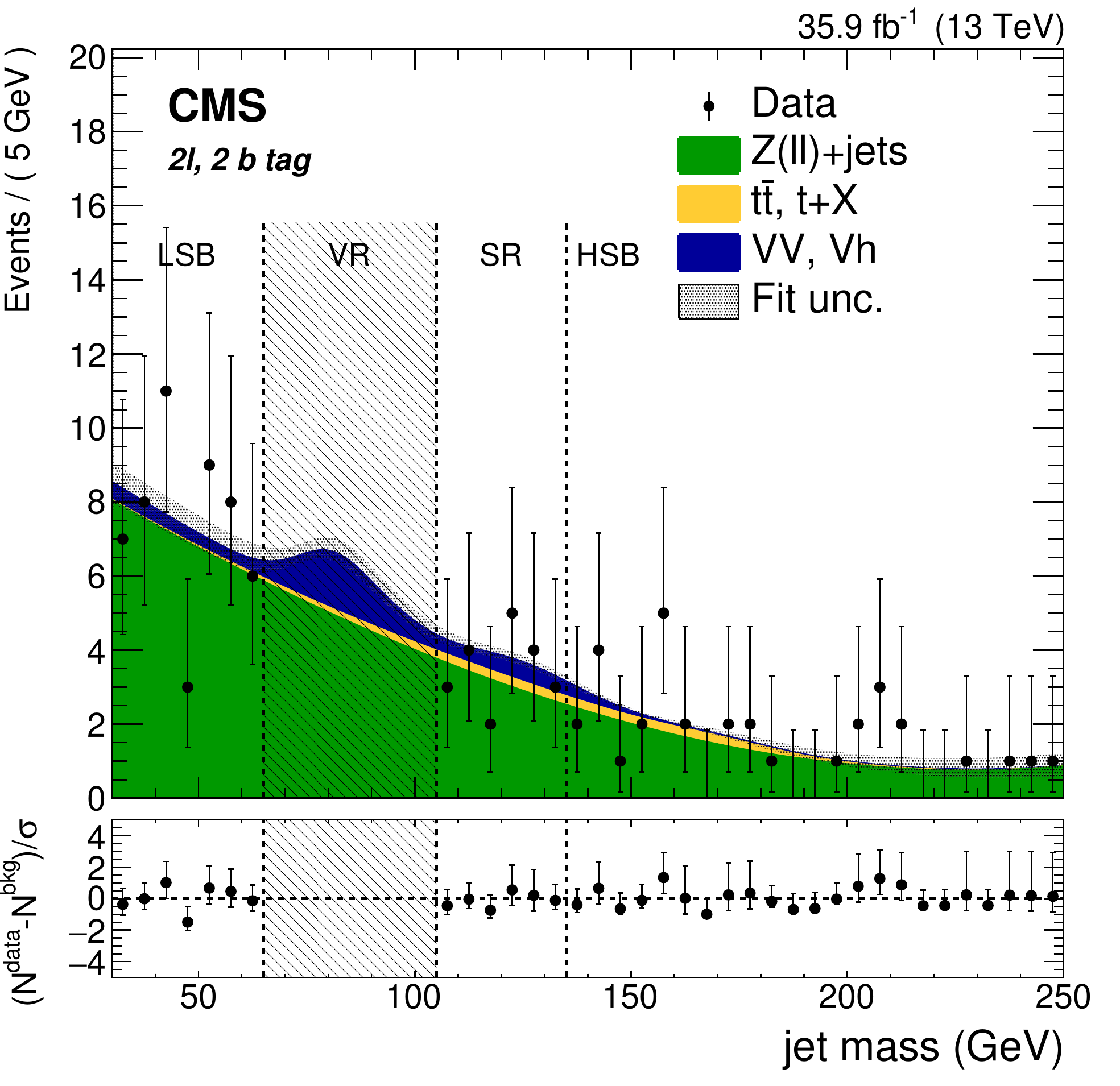}

  \caption{
    Soft-drop jet mass distribution of the leading AK8 jet in the $0\ell$ (upper), $1\ell$ (middle), and $2\ell$ (lower) categories, and separately for the 1 (\cmsLeft) and 2 (\cmsRight) {\cPqb}-tagged subjet selections. The electron and muon categories are merged together. The shaded band represents the uncertainty from the fit to data in the jet mass sidebands. The observed data are indicated by black markers. The dashed vertical lines separate the lower (LSB) and upper (HSB) sidebands, the signal region (SR), and the {\PW} and \Z bosons mass region (VR); the latter is not used in the fit to avoid biases from $\X\to\V\V$ signals. The bottom panels depict the pulls in each bin, $(N^\text{data}-N^\text{bkg})/\sigma$, where $\sigma$ is the statistical uncertainty in data, as given by the Garwood interval~\cite{doi:10.1093/biomet/28.3-4.437}.
    \label{fig:mJ} }
\end{figure*}

The background model, being the sum of the \Vjets, top quark, and diboson background components, is obtained by fitting the \mj spectrum in data in the two sideband (SB) regions, defined as the regions with \Ph jet mass in the ranges $30 < \mj < 65\GeV$ and $135 < \mj < 250\GeV$. The mass interval $65 < \mj < 105\GeV$ (VR), which contains vector boson merged decays, is excluded from the fit to avoid biases from a $\X\to\V\V$ potential signal; dedicated analyses in the $\V\V$ channel in the same final state are a subject of separate publications~\cite{Sirunyan:2018ivv,Sirunyan:2018iff,Sirunyan:2018hsl}.
In the fit, the normalization and shape parameters of the \Vjets background are free to vary, and those relative to the top quark and diboson backgrounds are determined from simulation. For each background, the expectation and the corresponding uncertainty are derived from the integral of the fitted shapes in the SR ($105 < \mj < 135\GeV$).
The procedure is repeated selecting an alternative function, consisting of the sum of an exponential and a Gaussian function, to model the \Vjets background distribution and estimate the bias induced by the choice of the \Vjets fit function. The difference between the integral in the SR obtained with the nominal and the alternative functions is considered as a systematic uncertainty.
The observed events in the SR are compatible within systematic and statistical uncertainties with the expected background events, and are reported separately for each category in Table~\ref{tab:BkgNorm}. The fits to the \mj distributions are shown in Fig.~\ref{fig:mJ}.

\begin{table*}[!htb]
  \topcaption{The expected and observed numbers of events in the signal regions depicted in Fig.~\ref{fig:mJ} are reported for the different event categories, along with the associated uncertainties from four sources: the \Vjets background uncertainty obtained from the correlated variation of the fit parameters used in the background model (fit); the uncertainty associated with the choice of fit function, estimated by comparing the nominal and an alternative function (alt); the statistical component of the uncertainties of the top quark scale factors, and the extrapolation uncertainty from the control regions to the SR; the \VV normalization uncertainties relative to the normalization and \mj modeling. A detailed description of the systematic uncertainties is provided in Section~\ref{sec:sys}.}\label{tab:BkgNorm}
  \centering
  \newcolumntype{x}{D{.}{.}{3.0}}
   \newcolumntype{y}{D{.}{.}{3.1}}
  \newcolumntype{z}{D{.}{.}{2.1}}
    \begin{tabular}{cc ccc cc}
      \hline
      \multicolumn{2}{c}{Category} & \Vjets ($\pm$fit) ($\pm$alt) & \ttbar, \ST & \VV, \VH & Bkg. sum & Observed \\
      \hline
      \multirow{5}{*}{1 {\cPqb} tag}
      & $0\ell$ & $694 \pm 17 \pm 4$ & $91 \pm 5$ & $34 \pm 8$ & $819 \pm 20$ & $849$ \\
      & $1\Pe$ & $603 \pm 37 \pm 72$ & $700 \pm 24$ & $38 \pm 10$ & $1369 \pm 85$ & $1389$ \\
      & $1\mu$ & $944 \pm 41 \pm 18$ & $835 \pm 28$ & $58 \pm 15$ & $1836 \pm 55$ & $1800$ \\
      & $2\Pe$ & $71 \pm 5 \pm 5$ & $2 \pm 1$ & $3 \pm 1$ & $76 \pm 7$ & $68$ \\
      & $2\mu$ & $78 \pm 5 \pm 5$ & $3 \pm 1$ & $4 \pm 1$ & $85 \pm 7$ & $95$ \\
      [\cmsTabSkip]
      \multirow{5}{*}{2 {\cPqb} tag}
      & $0\ell$ & $88 \pm 6 \pm 4$ & $17 \pm 2$ & $11 \pm 3$ & $116 \pm 8$ & $126$ \\
      & $1\Pe$ & $97 \pm 8 \pm 23$ & $146 \pm 7$ & $7 \pm 2$ & $249 \pm 25$ & $263$ \\
      & $1\mu$ & $131 \pm 9 \pm 13$ & $165 \pm 8$ & $10 \pm 3$ & $305 \pm 18$ & $316$ \\
      & $2\Pe$ & $8 \pm 1 \pm 1$ & $1 \pm 1$ & $1 \pm 1$ & $10 \pm 2$ & $7$ \\
      & $2\mu$ & $11 \pm 2 \pm 1$ & $1 \pm 1$ & $2 \pm 1$ & $13 \pm 2$ & $14$ \\
      \hline
    \end{tabular}
\end{table*}

\subsection{Background distribution}

The \mVH (or \mtVH) distribution of the \Vjets background is derived from data in the SB, and a transfer function $\alpha(\mVH)$ determined from simulation:
\begin{linenomath*}
\begin{equation}
\alpha(\mVH) = \frac{F_\mathrm{SR}^{\text{sim},\Vjets}(\mVH)}{F_\mathrm{SB}^{\text{sim},\Vjets}(\mVH)},
\end{equation}
\end{linenomath*}
where $F_\mathrm{SR}^{\text{sim},\Vjets}(\mVH)$, $F_\mathrm{SB}^{\text{sim},\Vjets}(\mVH)$ represent the probability density functions of the \Vjets background in the SR and SB regions, respectively. A two-parameter exponential $F(\mVH) = e^{~a~\mVH + b / \mVH}$ is chosen, using a simulated sample of \Vjets events. The background modelling is also performed using an alternative functional form $F(\mVH) = e^{- \mVH / (a + b~\mVH)}$. The resulting \Vjets prediction in the SR is found to be consistent within the uncertainties. 

\begin{figure*}[!hbtp]\centering
  \includegraphics[width=0.40\textwidth]{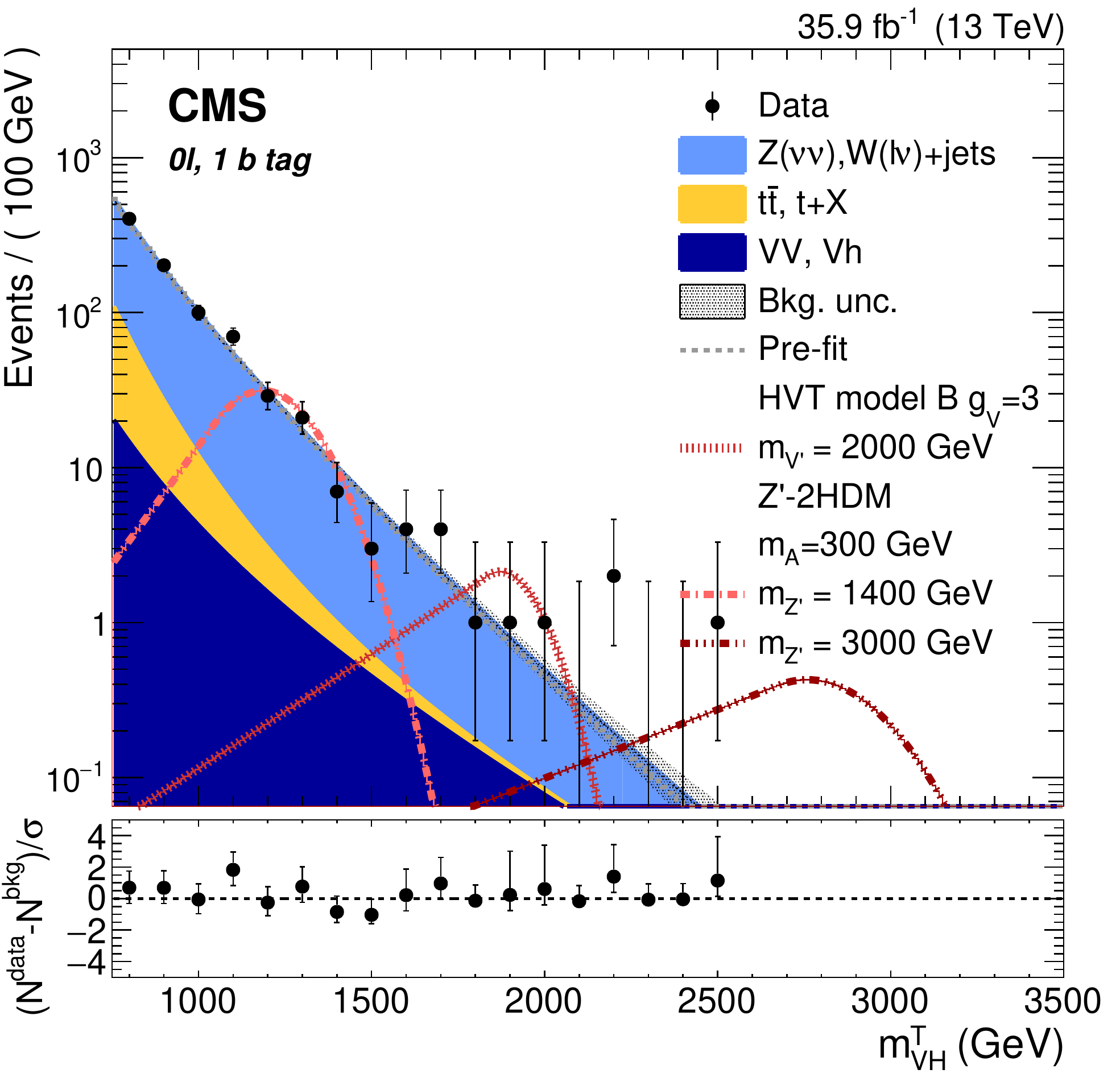}
  \includegraphics[width=0.40\textwidth]{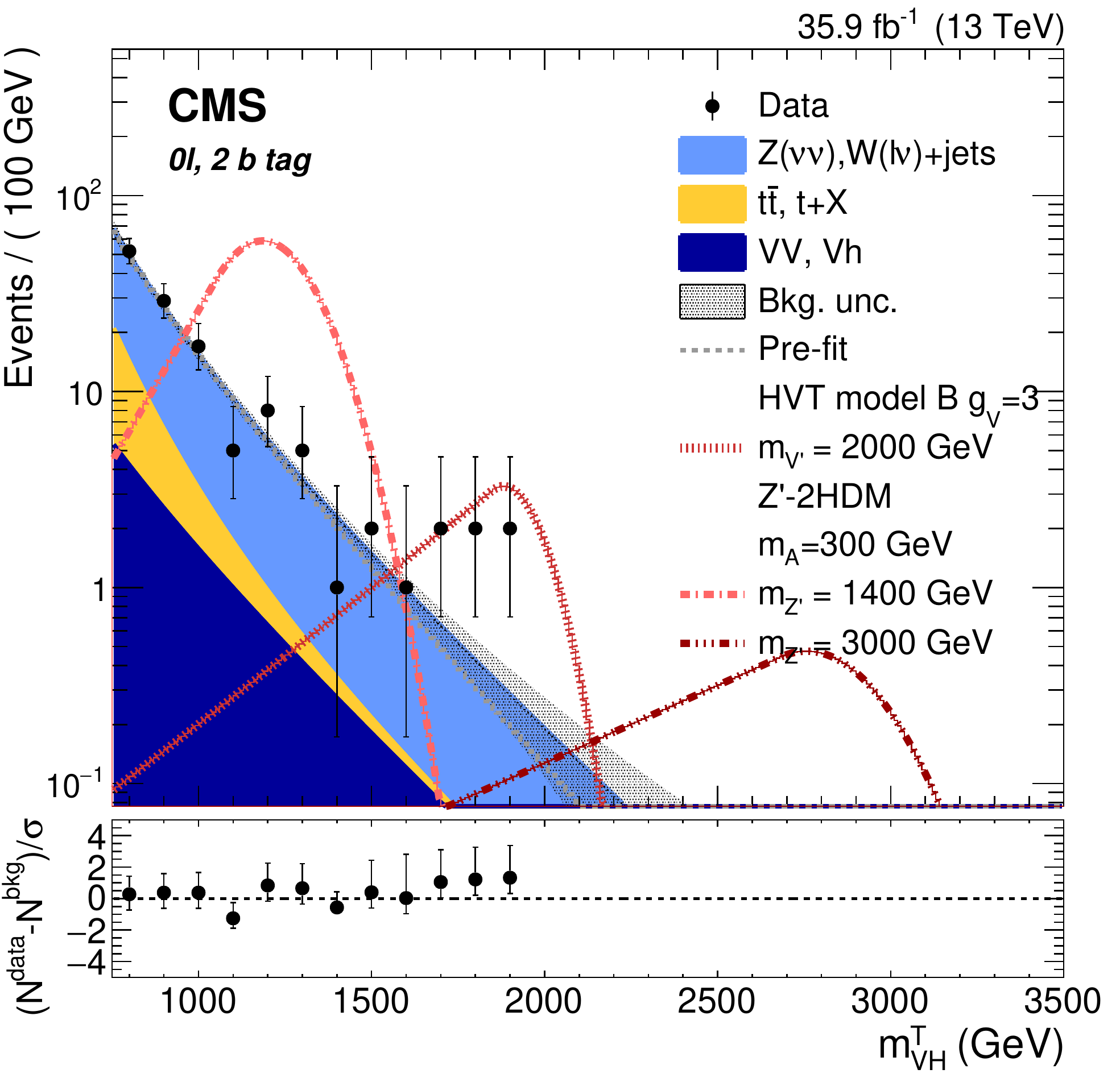}

  \includegraphics[width=0.40\textwidth]{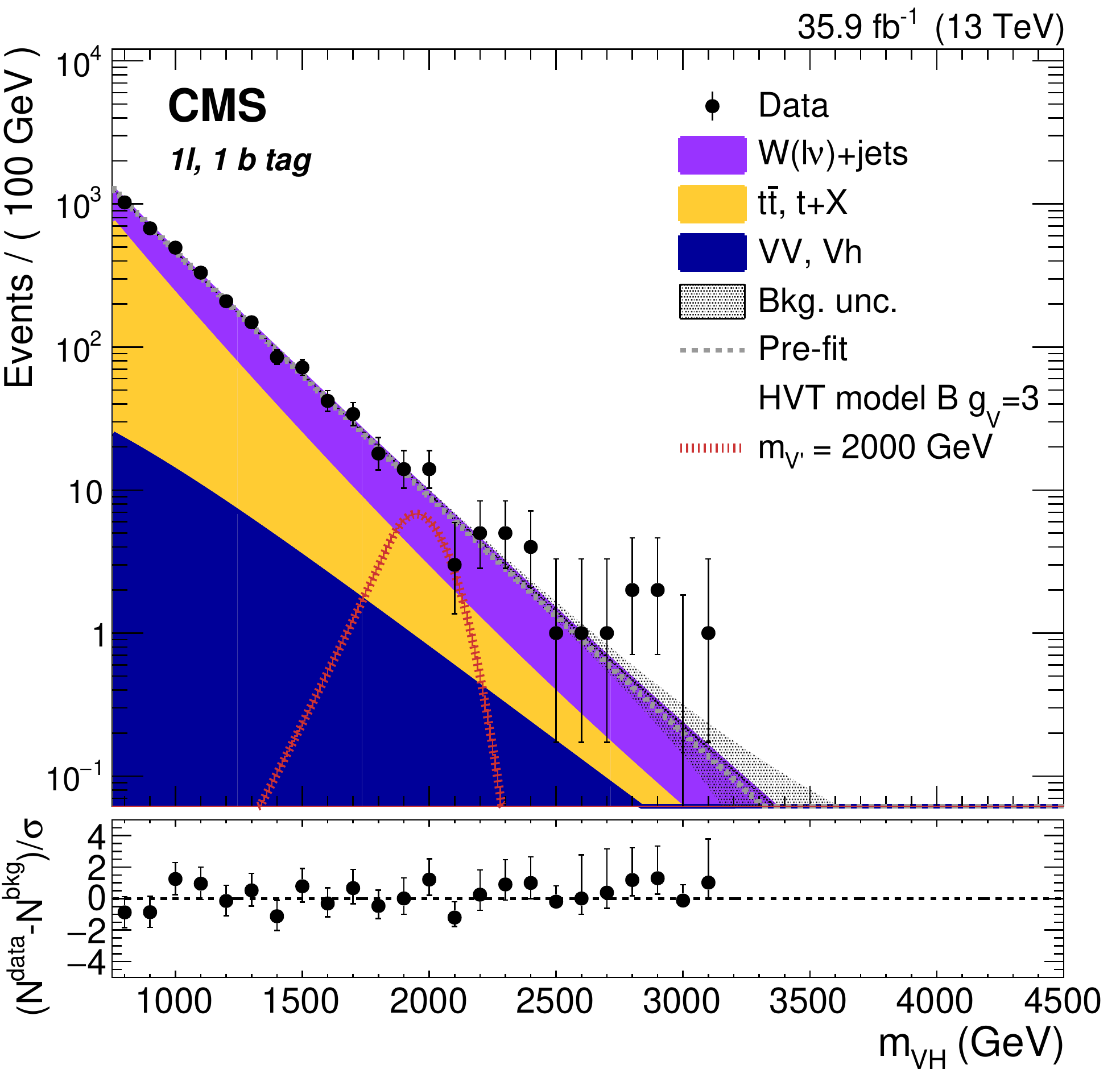}
  \includegraphics[width=0.40\textwidth]{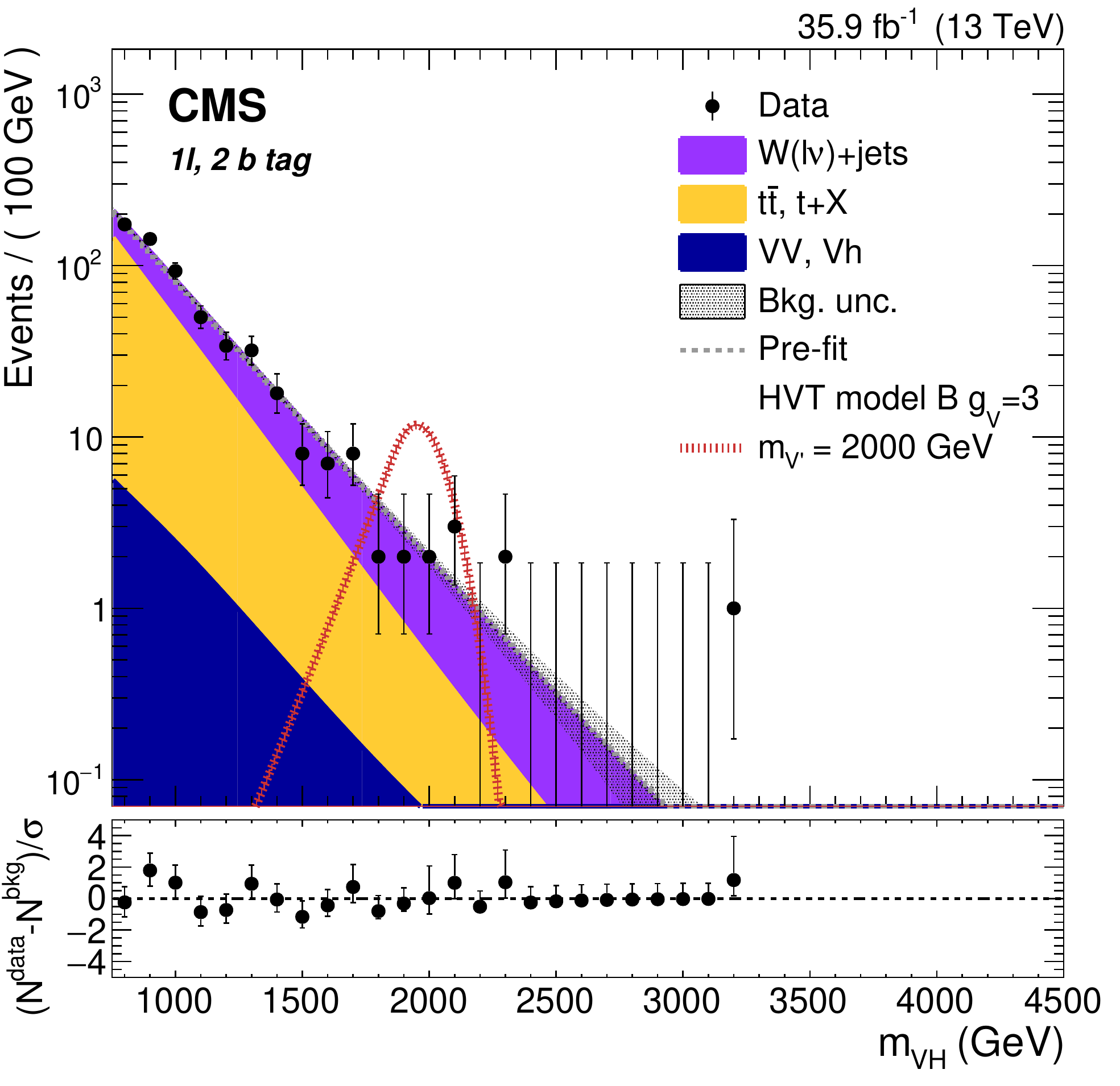}

  \includegraphics[width=0.40\textwidth]{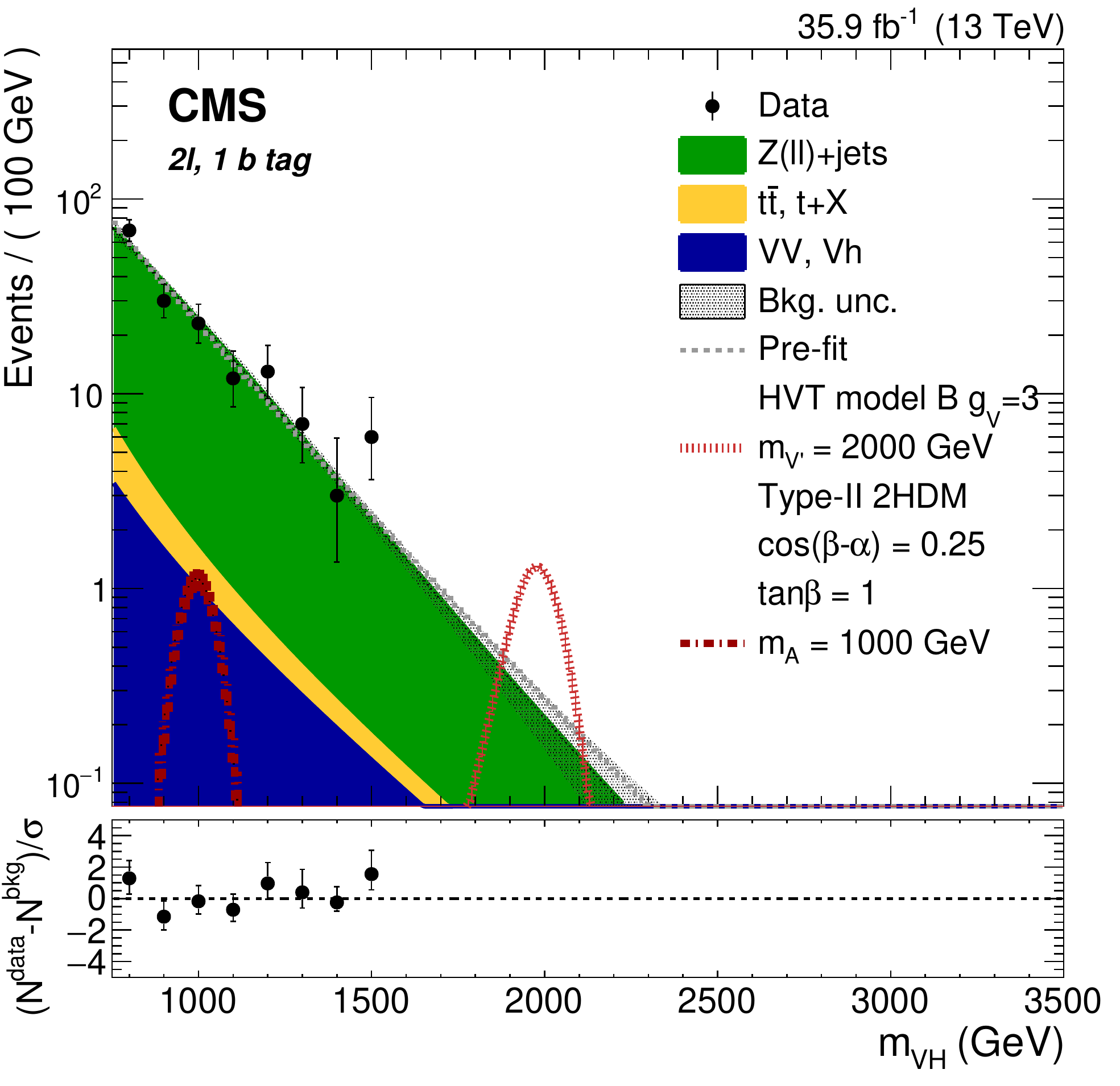}
  \includegraphics[width=0.40\textwidth]{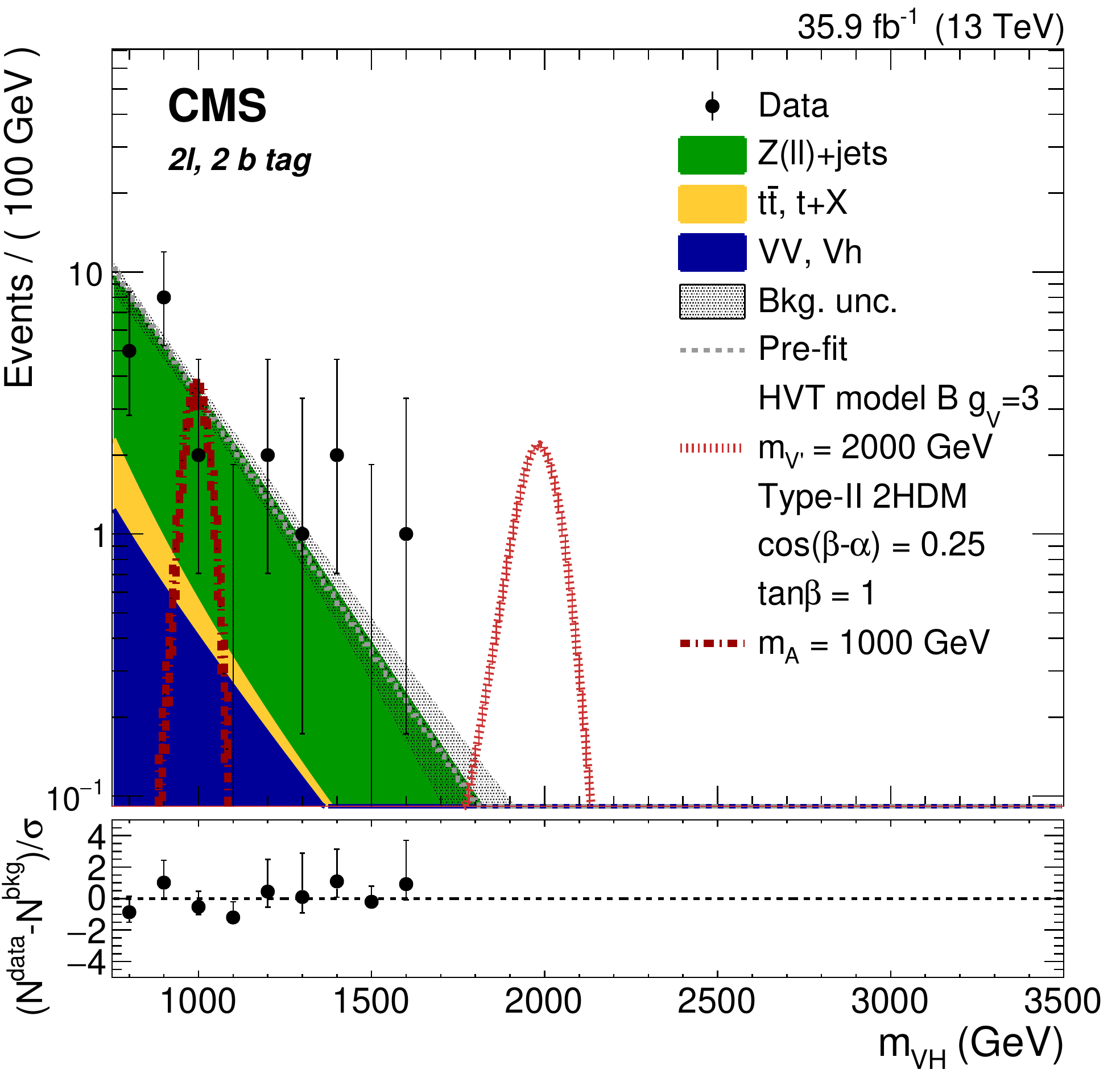}

  \caption{
    Resonance transverse mass \mtVH distributions in the $0\ell$ category (upper) and candidate mass \mVH in the $1\ell$ (middle), and $2\ell$ (lower) categories, and separately for the 1 (\cmsLeft) and 2 (\cmsRight) {\cPqb}-tagged subjet selections. Electron and muon categories are merged together. The expected background events are shown as filled areas, and the shaded band represents the total background uncertainty. The observed data are indicated by black markers, and the potential contribution of a resonance produced in the context of the HVT model B with $\gV=3$, or a \PZpr-2HDM signal with $\mA=300\GeV$, $m_\chi=100\GeV$, and $g_{\PZpr}=0.8$, are shown as dotted red lines.
    The bottom panels depict the pulls in each bin, $(N^\text{data}-N^\text{bkg})/\sigma$, where $\sigma$ is the statistical uncertainty in data, as given by the Garwood interval~\cite{doi:10.1093/biomet/28.3-4.437}.
    \label{fig:mX} }
\end{figure*}

The ratio $\alpha(\mVH)$ accounts for the correlations and the small kinematic differences involved in the interpolation from the SB regions to the SR, and is largely independent of the correlated uncertainties affecting the \mVH shape as they cancel out in the ratio.
The total background prediction in the SR $F_\mathrm{SR}^\text{pred}(\mVH)$ is extracted from data in the \mj SBs, after multiplying the obtained distribution by the $\alpha(\mVH)$ ratio:
\ifthenelse{\boolean{cms@external}}{
\begin{multline}
F_\mathrm{SR}^\text{pred}(\mVH) =  N_\mathrm{SB}^\text{\Vjets} F_\mathrm{SB}^\text{obs,\Vjets}(\mVH) \, \alpha(\mVH) \\+ N_\mathrm{SR}^\text{\ttbar} F_\mathrm{SR}^\text{sim,\ttbar}(\mVH) + N_\mathrm{SR}^\text{\VV} F_\mathrm{SR}^\text{sim,\VV}(\mVH),
\end{multline}
}{
\begin{linenomath*}
\begin{equation}
F_\mathrm{SR}^\text{pred}(\mVH) =  N_\mathrm{SB}^\text{\Vjets} F_\mathrm{SB}^\text{obs,\Vjets}(\mVH) \, \alpha(\mVH) + N_\mathrm{SR}^\text{\ttbar} F_\mathrm{SR}^\text{sim,\ttbar}(\mVH) + N_\mathrm{SR}^\text{\VV} F_\mathrm{SR}^\text{sim,\VV}(\mVH),
\end{equation}
\end{linenomath*}
}
where $F_\mathrm{SB}^\text{obs,\Vjets}(\mVH)$ is the probability distribution function obtained from a fit to data in the \mj SBs of the sum of the background components, and $F_\mathrm{SR}^\text{sim,\ttbar}(\mVH)$, and $F_\mathrm{SR}^\text{sim,\VV}(\mVH)$ are the shapes of the \ttbar and diboson components, respectively. The parameters $N_\mathrm{SB}^\text{\Vjets}$, $N_\mathrm{SR}^\text{\ttbar}$, and $N_\mathrm{SR}^\text{\VV}$ are instead determined from the fit to \mj, the top quark control regions, and simulated samples, respectively. The resulting background prediction is provided as input to the combined signal and background fit to the data in the SR discussed in Section~\ref{sec:res}. The data in the SR and the background predictions before and after the fit in the SR are shown in Fig.~\ref{fig:mX}.
The background estimation method is validated by splitting the lower \mj sideband into two regions with $30 < \mj < 50\GeV$ and $50 < \mj < 65\GeV$, and using the former interval to predict the background in the latter. The predicted yields and distributions are found to be compatible with the data.

\subsection{Signal modeling}

The signal \mVH or \mtVH mass shape is estimated from the simulated signal samples, parametrizing separately in each channel and signal hypotheses the signal distribution with a Gaussian peak, and a power law to model the lower mass tails.
The resolution of the reconstructed \mVH is given by the width of the Gaussian core for the $1\ell$ and $2\ell$ channels, and by the standard deviation of the \mtVH distribution in the $0\ell$ channel, and is found to be 10--16, 8--5, and 5--3\% of \mX in the $0\ell$, $1\ell$, and $2\ell$ channels, respectively, when going from low to high resonance masses.

\section{Systematic uncertainties}
\label{sec:sys}

The systematic uncertainty in the \Vjets and top quark background yields is dominated by the statistical uncertainty associated with the numbers of data events in the \mj SBs.
The uncertainties in the shapes of the \Vjets background and of the secondary backgrounds are estimated from the covariance matrix of the simultaneous fit of the \mVH or \mtVH distribution to data in the SBs and to simulated events in the SRs and SBs, and depend on the numbers of events in data and simulation in the corresponding regions.

The uncertainty in the top quark event yields can be attributed to the limited number of events in data and simulation in the respective control regions, as given in Table~\ref{tab:TopCR}. The uncertainties on the normalization associated with the event modeling and reconstruction are not considered in the SR, because the event yield of this background is taken from data. An additional uncertainty of 3\% is assigned to the extrapolation from the top quark control regions to the SR, and is estimated by inverting the {\cPqb} tag veto, for the $0\ell$ and $1\ell$ categories, or by changing the lepton flavor requirement, for the $2\ell$ category.
Minor contributions arise from the propagation of the uncertainties in the single top quark background and in the shape of the function modeling the \mj distributions of the \ttbar and \VV backgrounds.

Other sources of uncertainty affect both the normalization and shape of the simulated signal and the SM diboson background.
The uncertainties in the trigger efficiency and the electron, muon, and hadronic $\tau$ lepton reconstruction, identification, and isolation efficiencies are evaluated through studies of events with $\Z\to\ell\ell$ having the dilepton invariant mass around the \Z boson mass, and amount to approximately 2--5\% for the categories with charged leptons, and 1\% in the $0\ell$ categories.
The jet energy scale and resolution~\cite{Khachatryan:2016kdb} affect both shape and selection efficiencies, and are responsible for a 1\% variation in the numbers of background and signal events.
The jet mass scale and resolution uncertainties ranging from 1 to 6\% uncertainty for the SM diboson background, respectively, and to 11\% in the signal yields. The parton shower dependence of the jet mass scale and resolution is estimated using as an alternative the \mbox{\HERWIGpp} generator~\cite{Bellm:2015jjp,Bahr:2008pv}, based on which an additional uncertainty of 6\% is assigned.

The impact on the signal efficiency because of the {\cPqb} tagging systematic uncertainty~\cite{Sirunyan:2017ezt} depends on the \Ph jet \pt and thus on the mass of the resonance, and ranges from 2--5\% in the 1 {\cPqb} tag category to 3--7\% in the 2 {\cPqb} tag category.
The signal, \VV, and \ST background event yields and acceptances are affected by the choice of PDFs used by the event generators~\cite{Butterworth:2015oua} and the factorization and renormalization scale uncertainties. The former are derived with {\scshape SysCalc}~\cite{Kalogeropoulos:2018cke} according to the PDF4LHC recommendations~\cite{Butterworth:2015oua}, and the latter are estimated by varying the corresponding scales up and down by a factor of 2. The effect of these uncertainties is approximately 21\% for the \ttbar background, and for the signal is in the range 3--36\%, depending on the signal mass. The top quark background is also affected by the uncertainty in the \pt spectrum~\cite{Khachatryan:2016mnb}, which accounts for up to 14\% uncertainty propagated to the top quark background scale factors.
Additional systematic uncertainties affecting the event yield of backgrounds and signal, coming from pileup contributions, integrated luminosity~\cite{CMS:lumi}, the impact of jet energy scale and resolution on \ptmiss are also included in the analysis.

The fit parameters, normalization uncertainties, and \ttbar scale factors reported in Table~\ref{tab:TopCR} and Table~\ref{tab:BkgNorm} are statistically independent and are considered to be uncorrelated between the different categories. In contrast, the nuisance parameters relating to experimental effects or simulation uncertainties are assumed to be correlated.
A summary of the systematic uncertainties is given in Table~\ref{tab:Sys}.

\begin{table}[!htb]
  \centering
  \caption{Summary of systematic uncertainties for the backgrounds and signal samples. The entries labeled with $\checkmark$ are also propagated to the shapes of the distributions. The uncertainties marked with $\dagger$ have impact on the signal cross section. Uncertainties marked with $\ddagger$ only affect the top quark background scale factors.}
  \label{tab:Sys}
  \begin{tabular}{lccccc}
    \hline
                                    & shape & \Vjets & \ttbar, t+X & \VV, \VH & Signal \\
    \hline
    Bkg. normalization              & \NA & 2--15\% & \NA & \NA & \NA \\
    Top quark bkg. scale factors    & \NA & \NA & 2--17\% & \NA & \NA \\
    Jet energy scale & $\checkmark$ & \NA & \NA & 3\% & 1\% \\
    Jet energy resolution & $\checkmark$ & \NA & \NA & $<$1\% & $<$1\% \\
    Jet mass scale                  & \NA & \NA & \NA & 6\% & 1\% \\
    Jet mass resolution             & \NA & \NA & \NA & 6\% & 11\% \\
    Electron identification, isolation & \NA & \NA & 1--3\% & \multicolumn{2}{c}{1--4\%} \\
    Muon identification, isolation  & \NA & \NA & 1--3\% & \multicolumn{2}{c}{1--5\%} \\
    Lepton scale and resolution & $\checkmark$ & \NA & \NA & \NA & 1--5\% \\
    Hadronic $\tau$ veto            & \NA & \NA & \NA & \multicolumn{2}{c}{3\% ($0\ell$)} \\
    \ptmiss scale and resolution    & \NA & \NA & \NA & 1\% & 1\% \\
    Electron, muon, \ptmiss trigger & \NA & \NA & \NA & \multicolumn{2}{c}{3--4\%} \\
    \multirow{2}{*}{{\cPqb} tagging}      & \NA & \multirow{2}{*}{\NA} & 3\% ($0\ell$, $1\ell$) & 4\% (1{\cPqb}) & 2--5\% (1{\cPqb}) \\
                                    &   &   & 2--5\% $\ddagger$ & 5\% (2{\cPqb}) & 3--7\% (2{\cPqb}) \\
    Higgs boson jet                 & \NA & \NA & \NA & \NA & 6\% \\
    Top quark \pt                   & \NA & \NA & 6--14\% $\ddagger$ & \NA & \NA \\
    Pileup                          & \NA & \NA & $<$1\% & $<$1\% & $<$1\% \\
    Factorization and               & \multirow{2}{*}{\NA} & \multirow{2}{*}{\NA} & \multirow{2}{*}{21\% $\ddagger$} & \multirow{2}{*}{19\%} & \multirow{2}{*}{3--28\% $\dagger$} \\
    \quad renormalization scales \\
    PDF normalization               & \NA & \NA & 5\% $\ddagger$ & 5\% & 8--36\% $\dagger$ \\
    PDF acceptance                  & \NA & \NA & 2\% $\ddagger$ & $<$2\% & $<$1\% \\
    Luminosity                      & \NA & \NA & \NA & 2.5\% & 2.5\% \\
    \hline
  \end{tabular}
\end{table}

\section{Results and interpretation}
\label{sec:res}

The \mVH or \mtVH mass spectra in Fig.~\ref{fig:mX} are fit with a combined likelihood function. The results of the unbinned fit are interpreted in the context of different models.
Systematic uncertainties are treated as nuisance parameters and are profiled in the statistical interpretation~\cite{CLS1,CLS2,CMS-NOTE-2011-005}. The background-only hypothesis is tested against the $\X\to\VH$ signal in the ten categories.
The asymptotic modified frequentist method~\cite{Asymptotic} is used to determine limits at 95\% confidence level (\CL) on the product of the cross section for a heavy boson \X and the branching fractions for the decays $\X\to\VH$ and $\Phtobb$, denoted $\sigma(\X) \, \B(\X\to\VH) \, \B(\Phtobb)$.
The $0\ell$ and $2\ell$ categories are combined to provide upper limits for the case where \X is a heavy spin-1 vector singlet $\PZpr$ or a pseudoscalar boson \A; similarly, the $1\ell$ categories are combined to provide limits for the case where \X is a heavy $\PWpr$. The $0\ell$ categories are also used to place limits on the \PZpr-2HDM model. The largest excess, corresponding to a local significance of 2.3 standard deviations, is observed in the $0\ell$ category at $\mX \approx 2 \TeV$.
The uncertainties in the signal cross section (marked in Table~\ref{tab:Sys}) are not profiled in the fit when presenting the results as upper limits on the cross sections as a function of \mX, or as a function of $m_{\PZpr}$ and \mA in the \PZpr-2HDM model, and are included in the uncertainty band of the theoretical cross section line. When placing constraints on the HVT and 2HDM model parameters, the uncertainties are profiled in the fit.

\begin{figure}[!htb]\centering
    \includegraphics[width=0.495\textwidth]{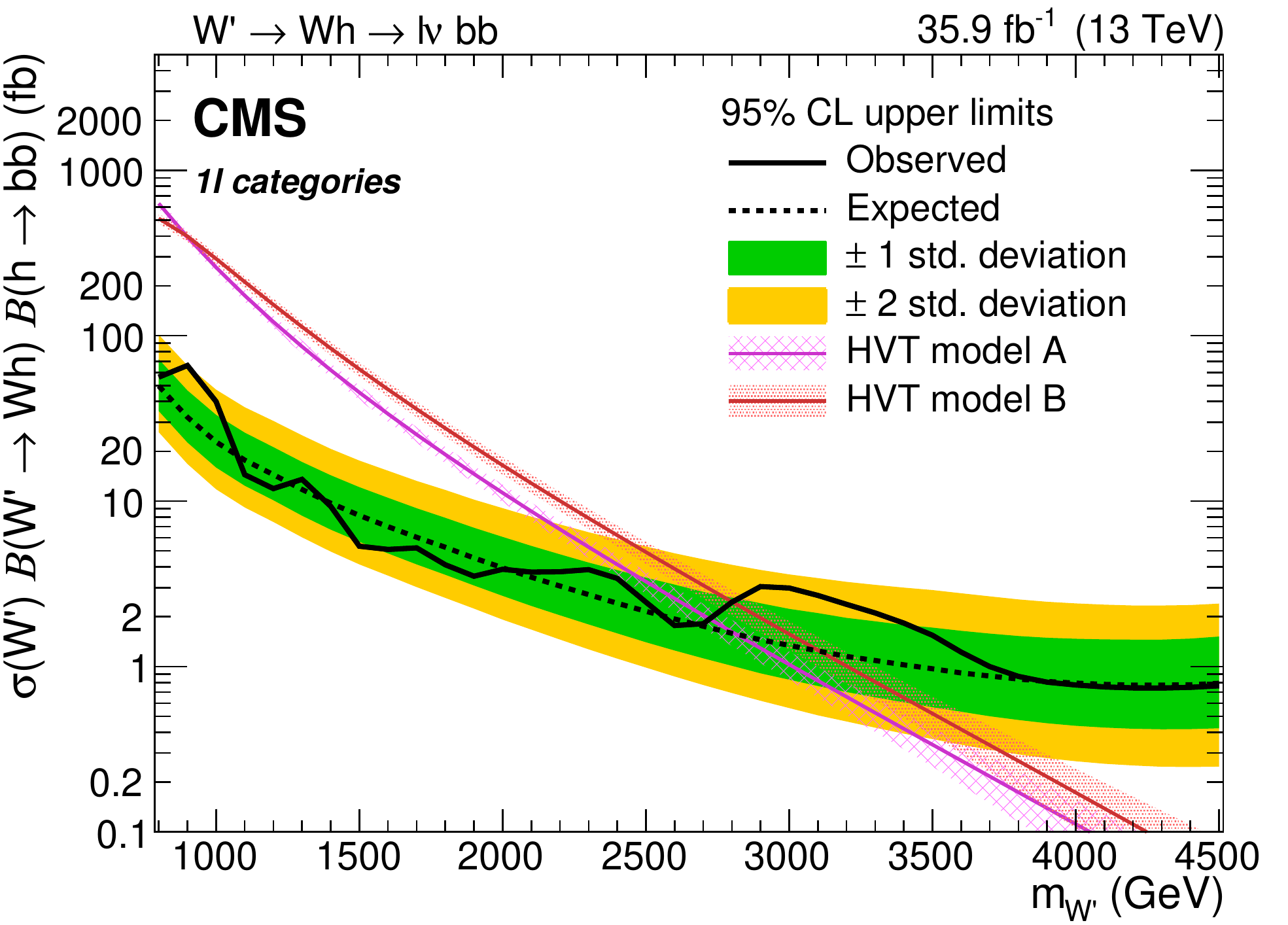}
    \includegraphics[width=0.495\textwidth]{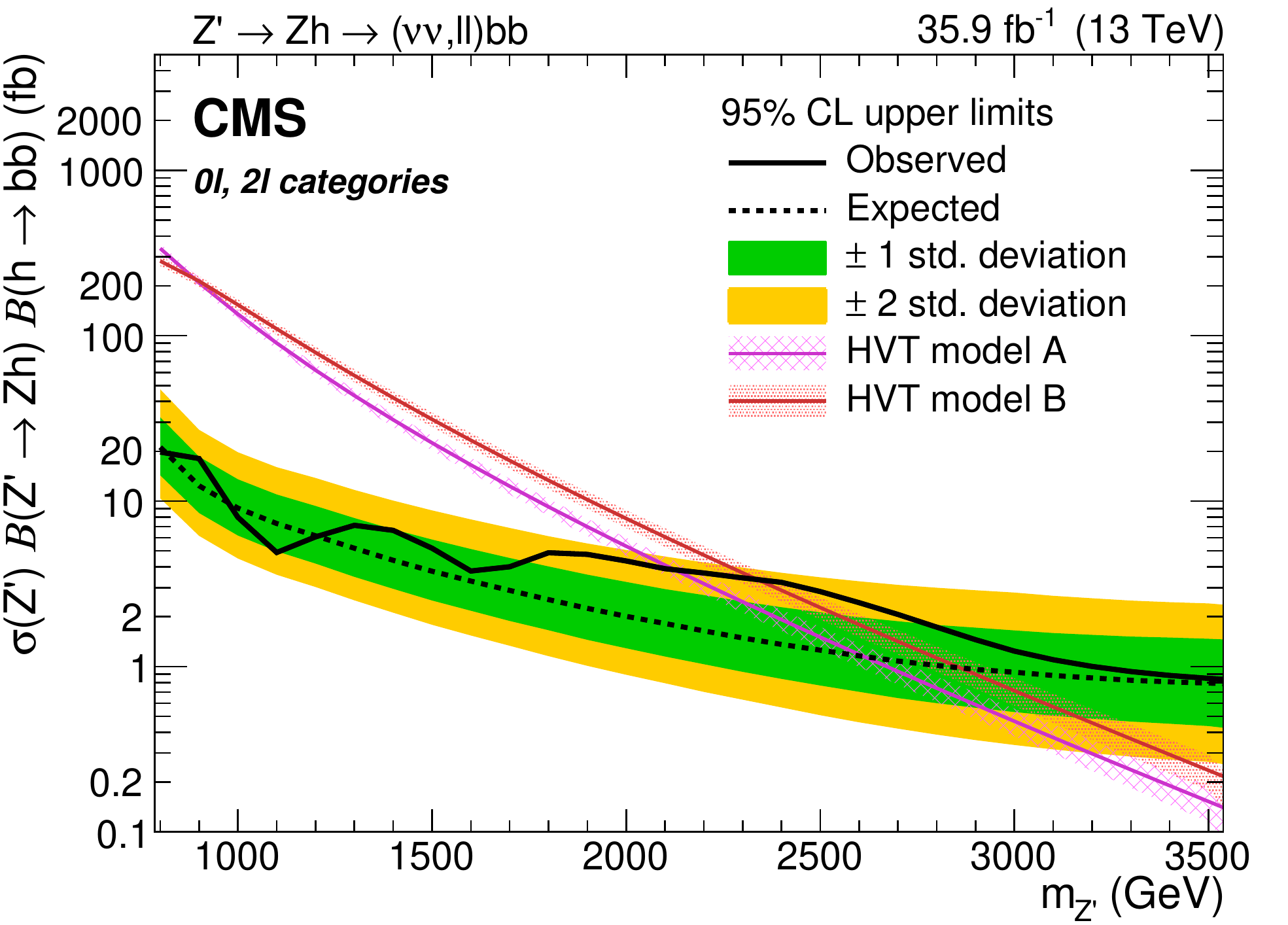}
    \caption{Observed and expected 95\% \CL upper limits on $\sigma(\PWpr) \, \B(\PWpr\to\PW\Ph) \, \B(\Phtobb)$ (\cmsLeft) and $\sigma(\PZpr) \, \B(\PZpr\to\Z\Ph) \, \B(\Phtobb)$ (\cmsRight) for various mass hypotheses of a single narrow spin-1 resonance. The inner green and outer yellow bands represent the ${\pm}1$ and ${\pm}2$ standard deviation (std.) variations on the expected limits. The solid curves and their shaded areas correspond to the product of the cross sections and the branching fractions predicted by the HVT models~A and B and the relative uncertainties.
    }
  \label{fig:limit}
\end{figure}

\begin{figure}[!htb]\centering
    \includegraphics[width=0.7\textwidth]{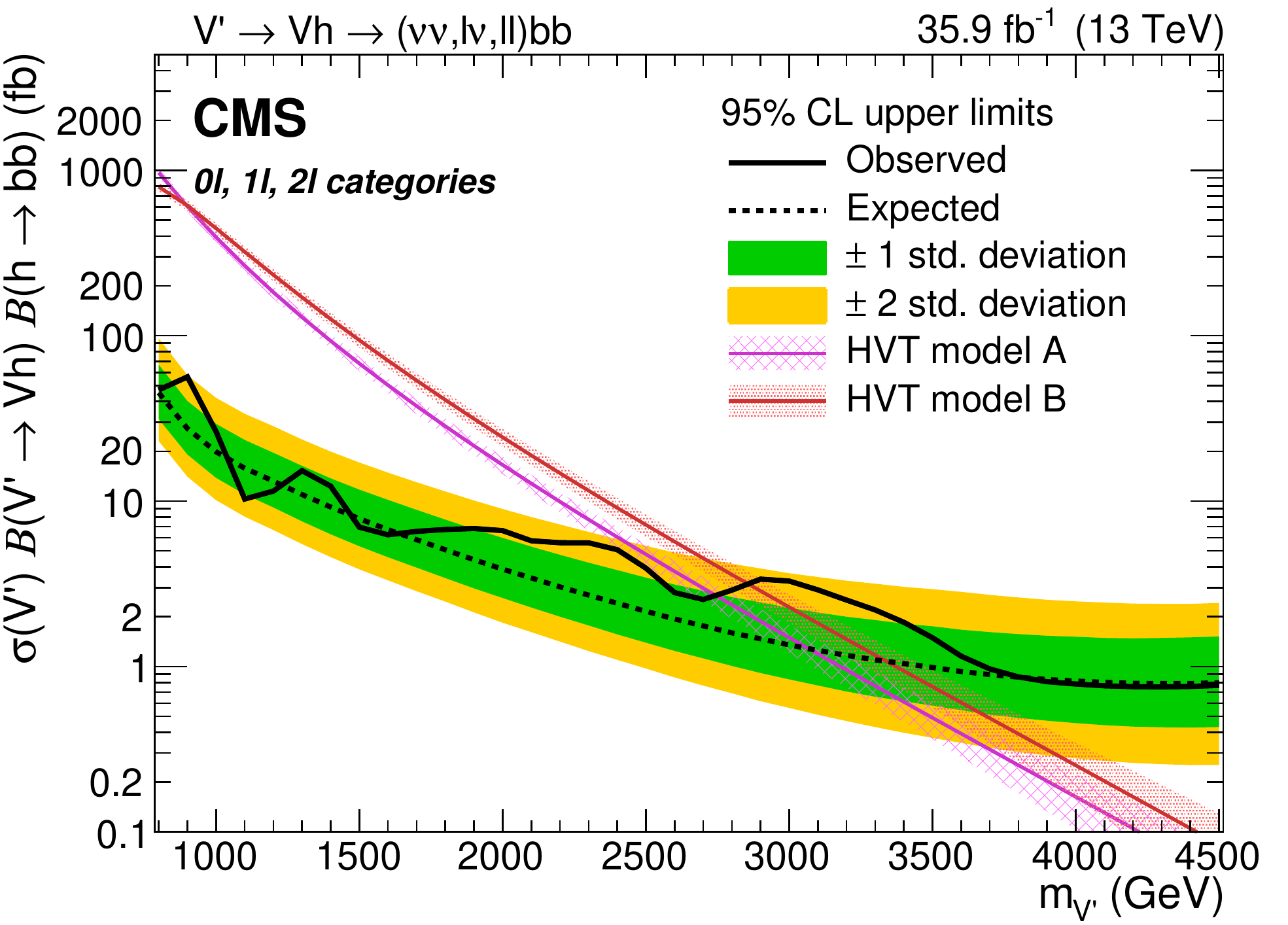}
    \caption{Observed and expected 95\% \CL upper limit on $\sigma(\X) \, \B(\X\to\VH) \, \B(\Phtobb)$ as a function of the HVT triplet mass, for the combination of all the considered channels. The inner green and outer yellow bands represent the ${\pm}1$ and ${\pm}2$ standard deviation (std.) variations on the expected limit. The solid curves and their shaded areas correspond to the cross sections predicted by the HVT models~A and B and the relative uncertainties.}
  \label{fig:xvh}
\end{figure}

The exclusion limits for the spin-1 singlet hypotheses (\PWpr or \PZpr) are shown in Fig.~\ref{fig:limit}. In the HVT model~B, a \PWpr and a \PZpr with mass lower than 2.8 and 2.3\TeV are excluded  at 95\% \CL, respectively. The HVT triplet hypothesis is tested by combining the $0\ell$, $1\ell$, and $2\ell$ categories and adding the \PZpr and \PWpr cross sections in Fig.~\ref{fig:xvh}, and taking into account the event migrations between signal categories if leptons do not pass the acceptance or analysis requirements. The predictions of the HVT models~A and B are superimposed on the exclusion limits, and a heavy triplet with $\mVpr < 2.8$ and $2.9\TeV$ is excluded in the HVT models~A and~B, respectively. These results are similar to those reported in the ATLAS search performed with the same final states in a comparable data set~\cite{Aaboud:2017cxo}.

\begin{figure}[!htb]\centering
    \includegraphics[width=0.7\textwidth]{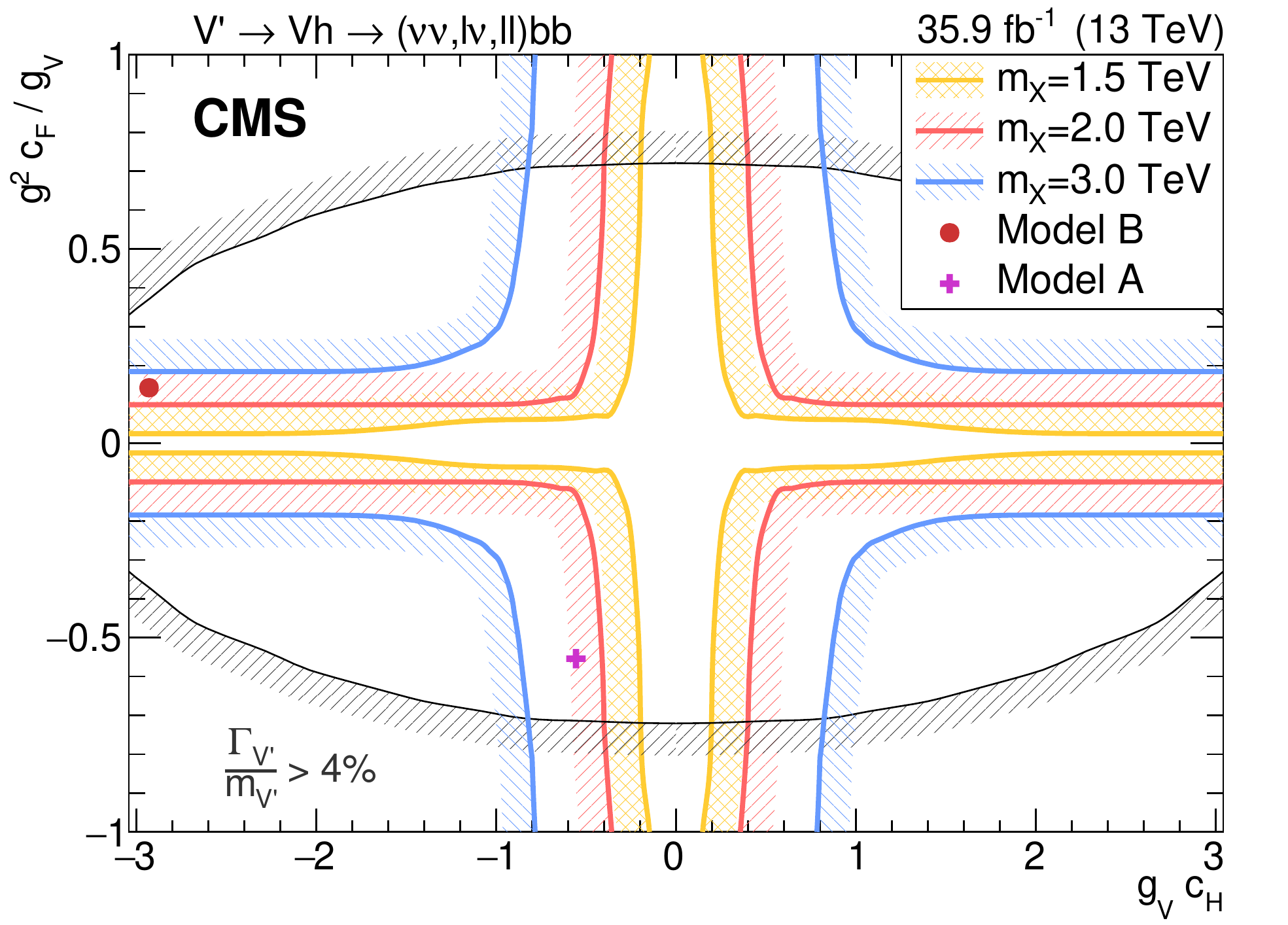}
    \caption{Observed exclusion limits in the HVT parameter plane $\left[ \gV \cH, \ g^2 \cF / \gV \right]$ for three different resonance masses (1.5, 2.0, and 3.0\TeV). The benchmark scenarios corresponding to HVT models~A and~B are represented by a purple cross and a red point. The areas bounded by the thin black contour lines correspond to the regions where the resonance natural width ($\Gamma_{\PVpr}$) is predicted to be larger than the typical experimental resolution (4\%), and the narrow-width approximation is no longer valid.}
  \label{fig:hvt}
\end{figure}

The exclusion limits on the resonance cross section shown in Fig.~\ref{fig:xvh} are also interpreted as a limit in the $\left[ \gV \cH, \ g^2 \cF / \gV \right]$ plane of the HVT parameters. The excluded region of the parameter space for narrow resonances obtained from the combination of all the considered channels is shown in Fig.~\ref{fig:hvt}. The fraction of the parameter space where the natural width of the resonances is larger than the average experimental resolution of 4\%, and the narrow-width approximation is not valid, is also indicated in Fig.~\ref{fig:hvt}.
The extent of the parameter space excluded significantly improves on the reach of the previous $\sqrt{s}=8$ and $13\TeV$ searches in the same final states~\cite{Khachatryan:2016yji,Khachatryan2017137,Aaboud:2017cxo}.

\begin{figure}[!htb]\centering
    \includegraphics[width=0.7\textwidth]{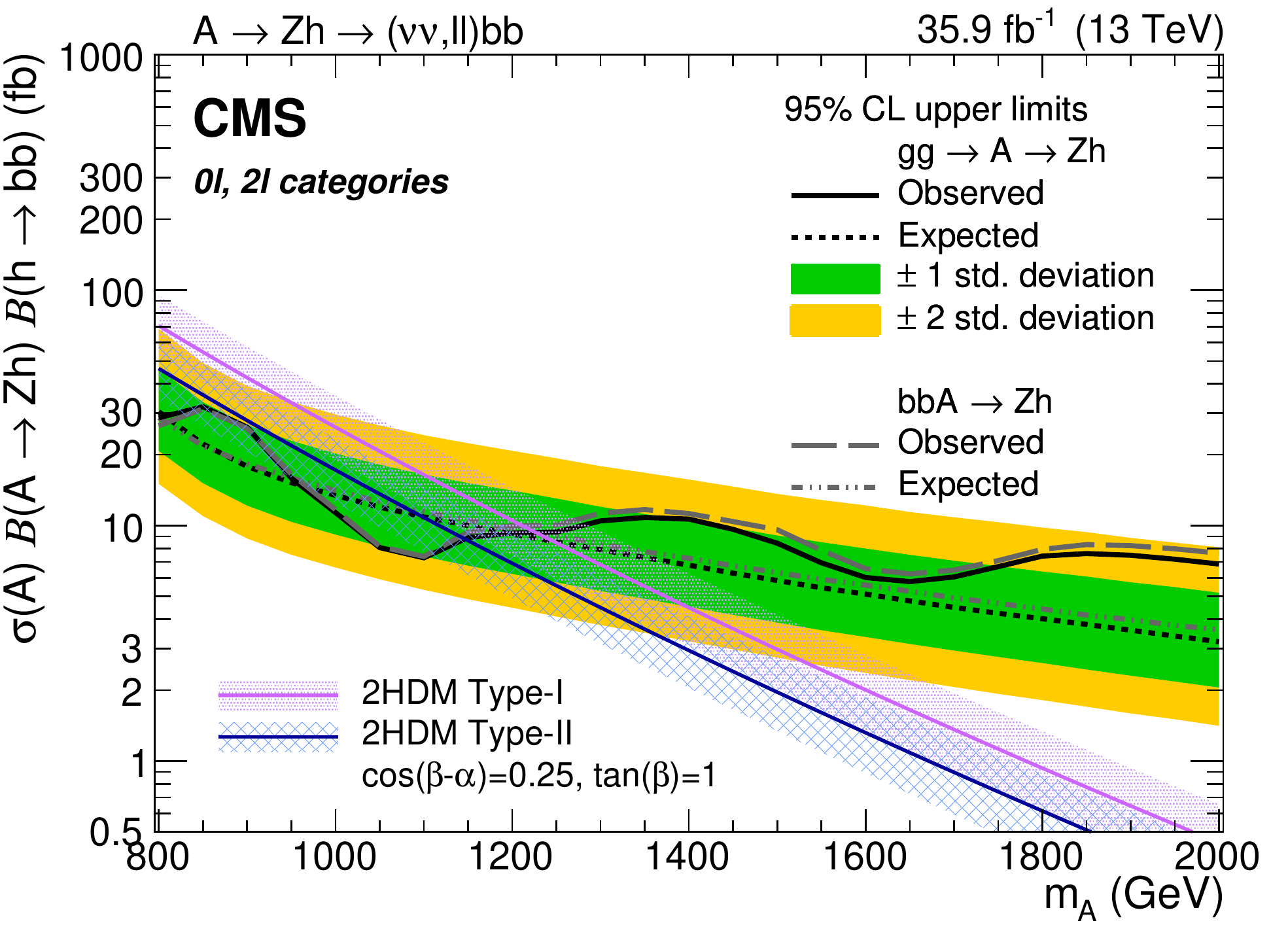}
    \caption{Observed and expected 95\% \CL upper limit on $\sigma(\A) \, \B(\A\to\Z\Ph) \, \B(\Ph\to\bbbar)$ as a function of \mA for the combination of the $0\ell$ and $2\ell$ channels. The inner green and outer yellow bands represent the ${\pm}1$ and ${\pm}2$ standard deviation (std.) variations on the expected limit. The solid line represent the exclusion for a spin-0 signal produced through gluon-gluon fusion, and dashed line represent the {\cPqb} quark associated production. The solid lines and their shaded areas represent the corresponding values predicted by the Type-I and Type-II 2HDM model fixing the parameters $\cosba=0.25$ and $\tan\beta=1$ parameters. In this scenario, the {\cPqb} quark associated production is negligible, and the \A boson is predominantly produced through gluon-gluon fusion.}
  \label{fig:azh}
\end{figure}

\begin{figure}[!htb]\centering
    \includegraphics[width=0.495\textwidth]{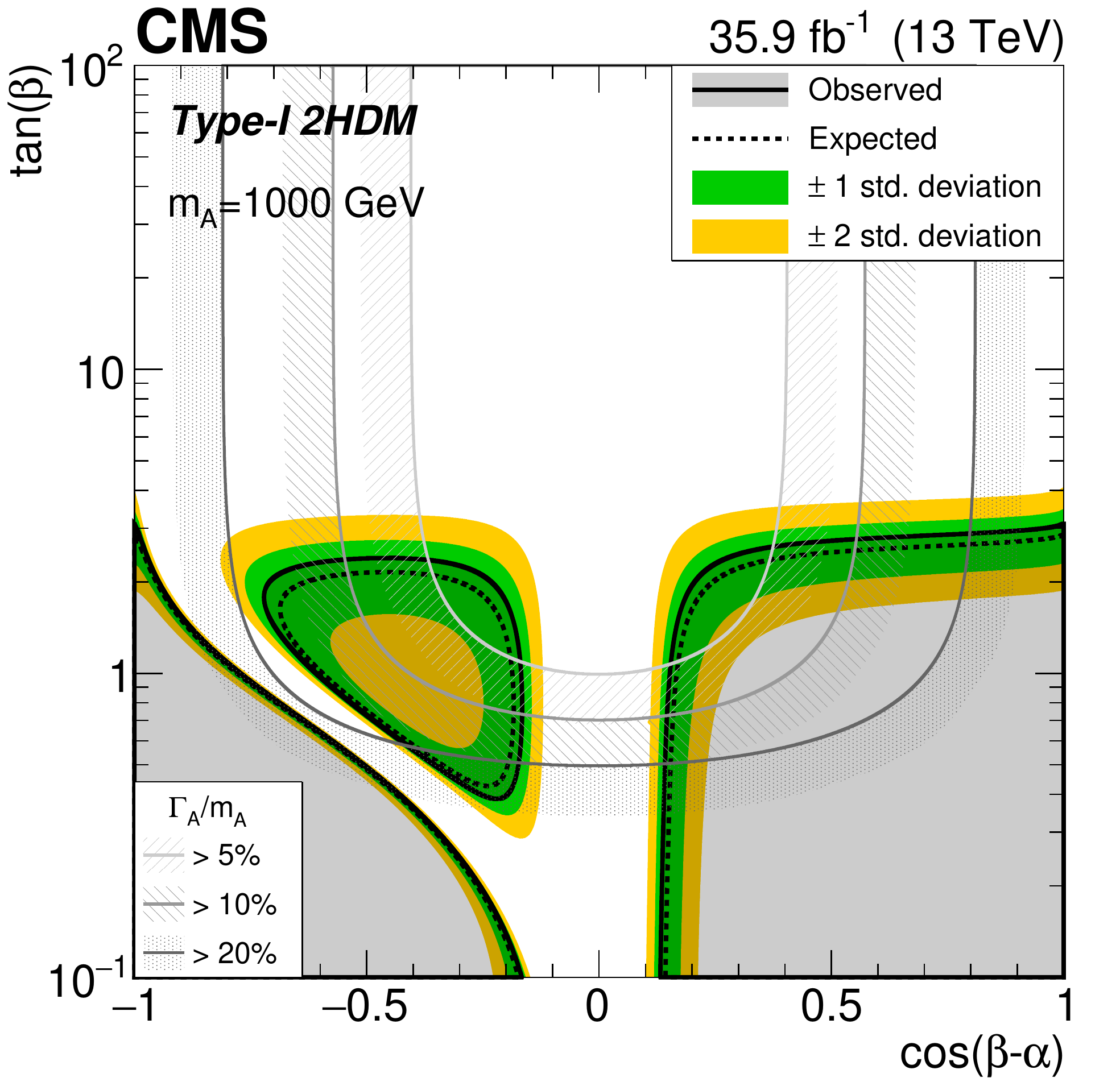}
    \includegraphics[width=0.495\textwidth]{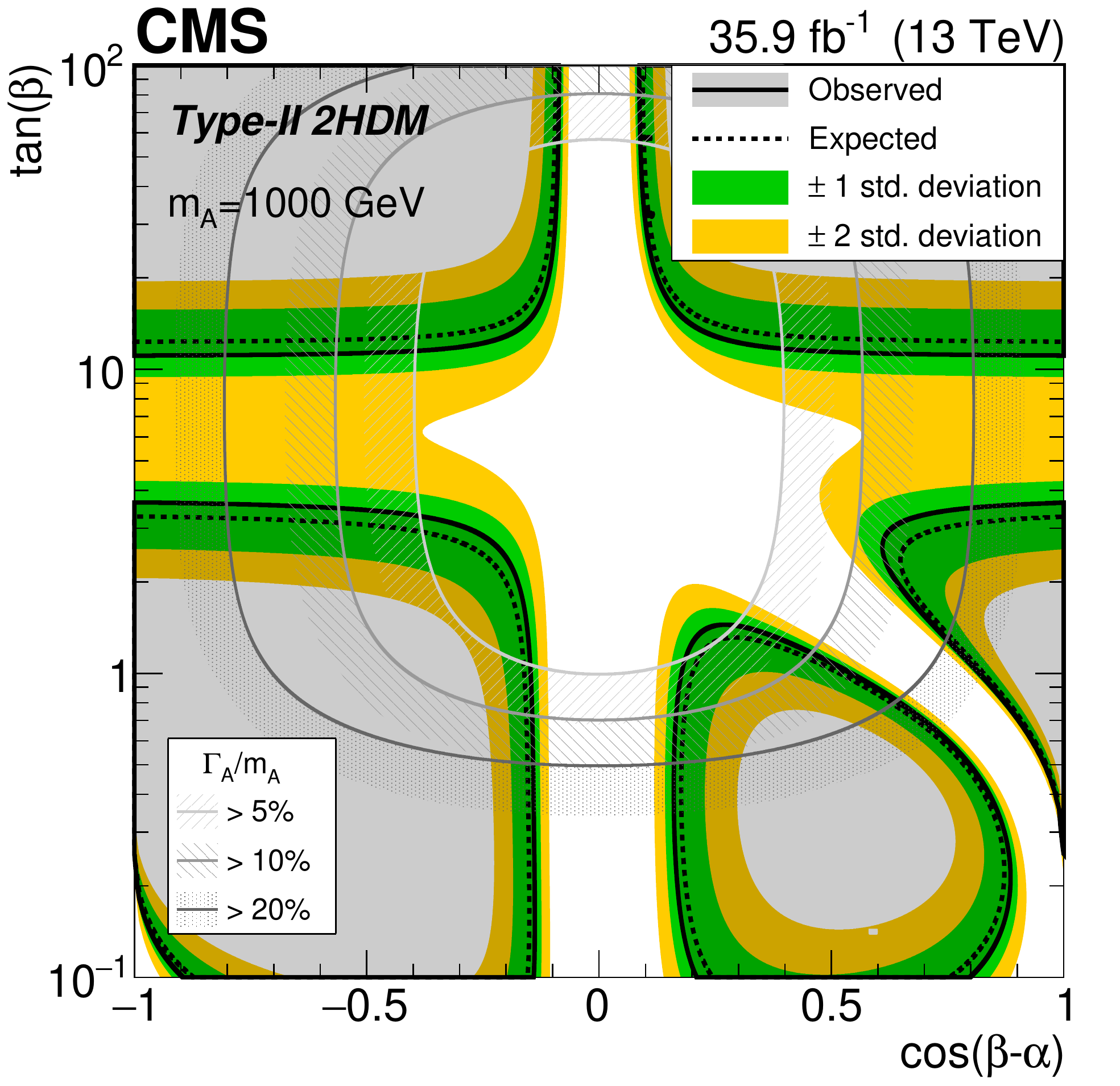}
    \caption{Observed and expected exclusion limit for Type-I (\cmsLeft) and Type-II (\cmsRight) 2HDM models in the [\tanb, \cosba] plane and assuming a fixed $\mA = 1\TeV$. The inner green and outer yellow bands represent the ${\pm}1$ and ${\pm}2$ standard deviation (std.) variations on the expected limit. The contour lines and associated shading identify regions with different resonance natural width (5, 10, and 20\% of the resonance mass).}
  \label{fig:2hdm}
\end{figure}

Figure~\ref{fig:azh} reports the exclusion limits as a function of the \A boson mass on the products of the \A boson cross section and the branching fraction $\B(\A\to\Z\Ph)$ and $\B(\Ph\to\bbbar)$, for production via gluon-gluon fusion or {\cPqb} quark associated production. The 2HDM cross sections and branching fractions are computed at NNLO with \textsc{2hdmc} 1.7.0~\cite{Eriksson2010189} and \textsc{SuShi} 1.6.1~\cite{Harlander20131605}, respectively. The parameters used for the models are: $\mh=125\GeV$, $\mH=m_{\PH^\pm}=\mA$, $m^2_{12}=\mA^2 \frac{\tan\beta}{1+\tan^2\beta}$ to break the discrete $Z_2$ symmetry as in the MSSM, and $\lambda_{6,7}=0$ to ensure CP conservation at tree level in the 2HDM Higgs sector~\cite{bib:Branco20121}.
In the scenario with $\cosba=0.25$ and $\tan\beta=1$, an \A boson with mass up to $1.15$ and $1.23\TeV$ is excluded in the Type-I and Type-II scenario of the 2HDM, respectively. The exclusion limits on the gluon-gluon fusion and {\cPqb} quark associated production are used to place constraints on the corresponding cross sections, which depend on the model parameters. Fig.~\ref{fig:2hdm} shows the excluded two-dimensional plane of the 2HDM parameters $[\cosba, \tanb]$, with fixed $m_\A=1.0\TeV$ in the range $0.1\le\tanb\le100$ and $-1\le\cosba\le1$, using the convention $0<\beta-\alpha<\pi$. These results extend the search for a 2HDM pseudoscalar boson \A up to 2\TeV, and provide comparable limits to the ATLAS search~\cite{Aaboud:2017cxo}.

\begin{figure}[!htb]\centering
    \includegraphics[width=0.7\textwidth]{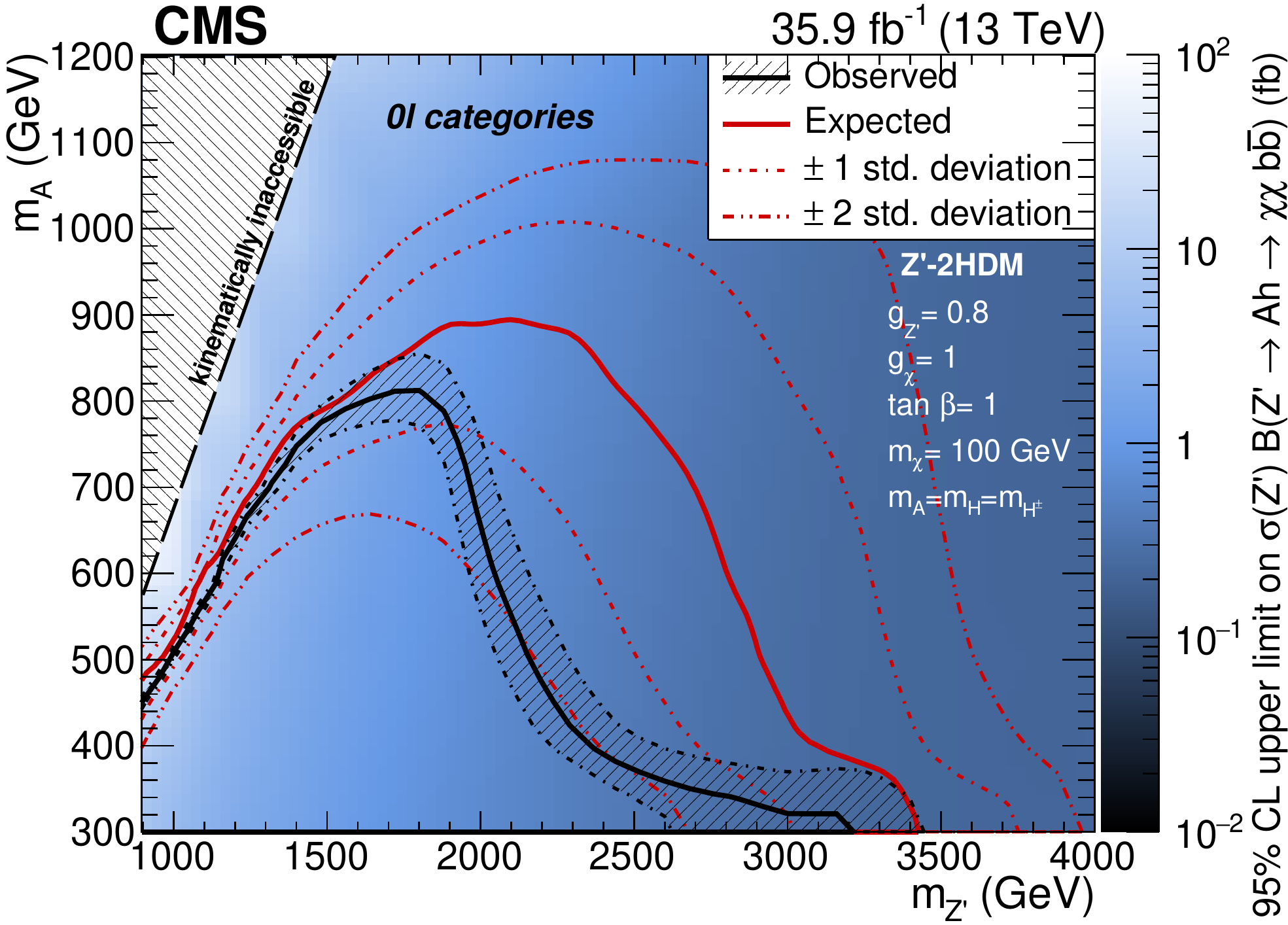}
    \caption{Observed and expected exclusions in the parameter plane [$m_{\PZpr}, m_\A$] at 95\% \CL. The excluded regions in the considered benchmark scenario ($g_{\PZpr} = 0.8$, $g_\chi = 1$, $\tan \beta = 1$, $m_\chi = 100\GeV$, and $\mA=\mH=\mHpm$) are represented by the areas below the curve. The hatched band relative to the observed limit represents the uncertainty on the signal cross section.}
  \label{fig:monoH_2D}
\end{figure}

The exclusion of the parameter space of the \PZpr-2HDM model is presented in Fig.~\ref{fig:monoH_2D} for the benchmark point with $g_{\PZpr} = 0.8$, $g_\chi = 1$, $m_\chi = 100\GeV$, and $\tan \beta = 1$. The branching fraction assumed for the \A boson decaying to DM particles is that predicted in the \PZpr-2HDM model, and SM branching fractions are assumed for the Higgs boson~\cite{Abercrombie:2015wmb}. The limits are presented for $m_{\PZpr}$ and $\mA$ parameter space in Fig.~\ref{fig:monoH_2D}. With the current data sample, $m_{\PZpr}$ up to 3.3\TeV and \mA up to $0.8\TeV$ are excluded, providing a more sensitive result compared to the ATLAS search performed on a similar data sample~\cite{Aaboud:2017yqz}, which excluded a $m_{\PZpr}< 2.5\TeV$ and $\mA<0.6\TeV$.

\section{Summary}
\label{sec:conclusions}

A search for resonances with masses between 800 and 4500\GeV, decaying to a standard model vector boson and the standard model Higgs boson, has been presented. The data sample was collected by the CMS experiment at $\sqrt{s}=13\TeV$, and corresponds to an integrated luminosity of 35.9\fbinv. The final states contain the leptonic decays of the vector bosons, in events with zero, exactly one, or two electrons or muons.
The \mVH or \mtVH mass spectra are used to fit for a localized excess consistent with a resonant signal, and no significant excess of events above the background predictions is observed.
Depending on the resonance mass, upper limits in the range 0.8--60\unit{fb} are set on the product of the cross sections and the branching fractions for the decay of the resonance into a Higgs boson and a vector boson, and with the subsequent decay of the Higgs boson into a pair of {\cPqb} quarks. Within the heavy vector triplet framework, vector bosons with a mass lower than 2.8 and 2.9\TeV are excluded for benchmark models~A and~B, respectively.
The results of this search also provide an exclusion in the two Higgs doublet model (2HDM) parameter space up to 2\TeV. A heavy pseudoscalar boson with mass lower than 1.1 and 1.2\TeV is excluded in the $\cosba=0.25$ and $\tan\beta=1$ scenario for Type-I and Type-II 2HDM, respectively.
A significant reduction of the allowed parameter space is also placed on the \PZpr-2HDM model that includes a dark matter candidate, excluding a \PZpr boson mass up to 3.3\TeV and a pseudoscalar boson \A with mass up to 0.8\TeV in the considered benchmark scenario. These are the most stringent limits placed on the \PZpr-2HDM model to date.

\begin{acknowledgments}
We congratulate our colleagues in the CERN accelerator departments for the excellent performance of the LHC and thank the technical and administrative staffs at CERN and at other CMS institutes for their contributions to the success of the CMS effort. In addition, we gratefully acknowledge the computing centers and personnel of the Worldwide LHC Computing Grid for delivering so effectively the computing infrastructure essential to our analyses. Finally, we acknowledge the enduring support for the construction and operation of the LHC and the CMS detector provided by the following funding agencies: BMWFW and FWF (Austria); FNRS and FWO (Belgium); CNPq, CAPES, FAPERJ, and FAPESP (Brazil); MES (Bulgaria); CERN; CAS, MoST, and NSFC (China); COLCIENCIAS (Colombia); MSES and CSF (Croatia); RPF (Cyprus); SENESCYT (Ecuador); MoER, ERC IUT, and ERDF (Estonia); Academy of Finland, MEC, and HIP (Finland); CEA and CNRS/IN2P3 (France); BMBF, DFG, and HGF (Germany); GSRT (Greece); NKFIA (Hungary); DAE and DST (India); IPM (Iran); SFI (Ireland); INFN (Italy); MSIP and NRF (Republic of Korea); LAS (Lithuania); MOE and UM (Malaysia); BUAP, CINVESTAV, CONACYT, LNS, SEP, and UASLP-FAI (Mexico); MBIE (New Zealand); PAEC (Pakistan); MSHE and NSC (Poland); FCT (Portugal); JINR (Dubna); MON, RosAtom, RAS and RFBR (Russia); MESTD (Serbia); SEIDI, CPAN, PCTI and FEDER (Spain); Swiss Funding Agencies (Switzerland); MST (Taipei); ThEPCenter, IPST, STAR, and NSTDA (Thailand); TUBITAK and TAEK (Turkey); NASU and SFFR (Ukraine); STFC (United Kingdom); DOE and NSF (USA).

\hyphenation{Rachada-pisek} Individuals have received support from the Marie-Curie program and the European Research Council and Horizon 2020 Grant, contract No. 675440 (European Union); the Leventis Foundation; the A. P. Sloan Foundation; the Alexander von Humboldt Foundation; the Belgian Federal Science Policy Office; the Fonds pour la Formation \`a la Recherche dans l'Industrie et dans l'Agriculture (FRIA-Belgium); the Agentschap voor Innovatie door Wetenschap en Technologie (IWT-Belgium); the F.R.S.-FNRS and FWO (Belgium) under the ``Excellence of Science - EOS" - be.h project n. 30820817; the Ministry of Education, Youth and Sports (MEYS) of the Czech Republic; the Lend\"ulet (``Momentum") Program and the J\'anos Bolyai Research Scholarship of the Hungarian Academy of Sciences, the New National Excellence Program \'UNKP, the NKFIA research grants 123842, 123959, 124845, 124850 and 125105 (Hungary); the Council of Science and Industrial Research, India; the HOMING PLUS program of the Foundation for Polish Science, cofinanced from European Union, Regional Development Fund, the Mobility Plus program of the Ministry of Science and Higher Education, the National Science Center (Poland), contracts Harmonia 2014/14/M/ST2/00428, Opus 2014/13/B/ST2/02543, 2014/15/B/ST2/03998, and 2015/19/B/ST2/02861, Sonata-bis 2012/07/E/ST2/01406; the National Priorities Research Program by Qatar National Research Fund; the Programa Estatal de Fomento de la Investigaci{\'o}n Cient{\'i}fica y T{\'e}cnica de Excelencia Mar\'{\i}a de Maeztu, grant MDM-2015-0509 and the Programa Severo Ochoa del Principado de Asturias; the Thalis and Aristeia programs cofinanced by EU-ESF and the Greek NSRF; the Rachadapisek Sompot Fund for Postdoctoral Fellowship, Chulalongkorn University and the Chulalongkorn Academic into Its 2nd Century Project Advancement Project (Thailand); the Welch Foundation, contract C-1845; and the Weston Havens Foundation (USA).

\end{acknowledgments}
\bibliography{auto_generated}
\cleardoublepage \appendix\section{The CMS Collaboration \label{app:collab}}\begin{sloppypar}\hyphenpenalty=5000\widowpenalty=500\clubpenalty=5000\vskip\cmsinstskip
\textbf{Yerevan Physics Institute, Yerevan, Armenia}\\*[0pt]
A.M.~Sirunyan, A.~Tumasyan
\vskip\cmsinstskip
\textbf{Institut f\"{u}r Hochenergiephysik, Wien, Austria}\\*[0pt]
W.~Adam, F.~Ambrogi, E.~Asilar, T.~Bergauer, J.~Brandstetter, E.~Brondolin, M.~Dragicevic, J.~Er\"{o}, A.~Escalante~Del~Valle, M.~Flechl, R.~Fr\"{u}hwirth\cmsAuthorMark{1}, V.M.~Ghete, J.~Hrubec, M.~Jeitler\cmsAuthorMark{1}, N.~Krammer, I.~Kr\"{a}tschmer, D.~Liko, T.~Madlener, I.~Mikulec, N.~Rad, H.~Rohringer, J.~Schieck\cmsAuthorMark{1}, R.~Sch\"{o}fbeck, M.~Spanring, D.~Spitzbart, A.~Taurok, W.~Waltenberger, J.~Wittmann, C.-E.~Wulz\cmsAuthorMark{1}, M.~Zarucki
% ...
\vskip\cmsinstskip
% ...
1:  Also at Vienna University of Technology, Vienna, Austria\\
% ...
\end{sloppypar}
\end{document}